\documentclass[twocolumn,tighten,times]{aastex61}

\newcommand\aastex{AAS\TeX}

\newcommand{\rA}{\mathrm{\AA}}
\newcommand{\msun}{\mathrm{M}_{\sun}}
\newcommand{\logm}{\mathrm{logM/\msun}}

\newcommand{\ha}{H$\alpha$}
\newcommand{\hb}{H$\beta$}
\newcommand{\hg}{H$\gamma$}
\newcommand{\hd}{H$\delta$}
\newcommand{\dfn}{D$4000_n$}
\newcommand{\hda}{\hd$_\mathrm{A}$}

\newcommand{\nuse}{1442}
\newcommand{\hblim}{$44.9\times10^{-19}\ \mathrm{ergs\ cm^{-2}\ s^{-1}}$}
\newcommand{\sfrlima}{$0.23\ \msun\ \mathrm{yr^{-1}}$}
\newcommand{\sfrlimb}{$0.47\ \msun\ \mathrm{yr^{-1}}$}
\newcommand{\sfrlimc}{$0.72\ \msun\ \mathrm{yr^{-1}}$}
\newcommand{\medianz}{$z_{\mathrm{spec}}=0.697$}
\newcommand{\sfrlimfinal}{$2.2\ \msun\ \mathrm{yr^{-1}}$}

\newcommand{\totn}{1988}
\newcommand{\prims}{1550}
\newcommand{\fills}{438}

\newcommand\suppress[1]{}

\shorttitle{\aastex\ LEGA-C Data Release II}
\shortauthors{Straatman et al.}

\begin{document}

\title{The Large Early Galaxy Astrophysics Census (LEGA-C) Data Release II: dynamical and stellar population properties of $z\lesssim1$ galaxies in the COSMOS field}

\correspondingauthor{Caroline M. S. Straatman}
\email{Caroline.Straatman@Ugent.be}

\author[0000-0001-5937-4590]{Caroline M. S. Straatman}
\affil{Sterrenkundig Observatorium, Universiteit Gent, Krijgslaan 281 S9, 9000 Gent, Belgium}

\author[0000-0002-5027-0135]{Arjen van der Wel}
\affil{Sterrenkundig Observatorium, Universiteit Gent, Krijgslaan 281 S9, 9000 Gent, Belgium}
\affil{Max-Planck Institut f\"ur Astronomie, K\"onigstuhl 17, D-69117, Heidelberg, Germany}

\author{Rachel Bezanson}
\affil{University of Pittsburgh, Department of Physics and Astronomy, 100 Allen Hall, 3941 O’Hara St, Pittsburgh PA 15260, USA}

\author{Camilla Pacifici}
\affil{Space Telescope Science Institute, 3700 San Martin Drive, Baltimore, MD 21218, USA}

\author[0000-0002-9656-1800]{Anna Gallazzi}
\affil{INAF-Osservatorio Astrofisico di Arcetri, Largo Enrico, Fermi 5, I-50125 Firenze, Italy}

\author{Po-Feng Wu}
\affil{Max-Planck Institut f\"ur Astronomie, K\"onigstuhl 17, D-69117, Heidelberg, Germany}

\author{Kai Noeske}
\affil{Experimenta Heilbronn, Kranenstraße 14, 74072, Heilbronn, Germany}

\author{Ivana Bari\v si\'c}
\affil{Max-Planck Institut f\"ur Astronomie, K\"onigstuhl 17, D-69117, Heidelberg, Germany}

\author{Eric F. Bell}
\affil{Department of Astronomy, University of Michigan, 1085 S. University Ave., Ann Arbor, MI 48109, USA}

\author{Gabriel B. Brammer}
\affil{Space Telescope Science Institute, 3700 San Martin Drive, Baltimore, MD 21218, USA}
\affil{Cosmic Dawn Center, Niels Bohr Institute, University of Copenhagen, Juliane Maries Vej 30, DK-2100 Copenhagen Ø, Denmark}

\author{Jo\~ao Calhau}
\affil{Department of Physics, Lancaster University, Lancaster LA14YB, UK}

\author{Priscilla Chauke}
\affil{Max-Planck Institut f\"ur Astronomie, K\"onigstuhl 17, D-69117, Heidelberg, Germany}

\author{Marijn Franx}
\affil{Leiden Observatory, Leiden Unversity, P. O. Box 9513, NL-2300 AA Leiden, The Netherlands}

\author{Josha van Houdt}
\affil{Max-Planck Institut f\"ur Astronomie, K\"onigstuhl 17, D-69117, Heidelberg, Germany}

\author{Ivo Labb\'e}
\affil{Centre for Astrophysics and Supercomputing, Swinburne University, Hawthorn, VIC 3122, Australia}

\author{Michael V. Maseda}
\affil{Leiden Observatory, Leiden Unversity, P. O. Box 9513, NL-2300 AA Leiden, The Netherlands}

\author{Juan C. Mu\~noz-Mateos}
\affil{European Southern Observatory, Alonso de C\'ordova 3107, Casilla 19001, Vitacura, Santiago, Chile}

\author{Adam Muzzin}
\affil{Department of Physics and Astronomy, York University, 4700 Keele St., Toronto, Ontario, MJ3 1P3, Canada}

\author[0000-0003-2552-0021]{Jesse van de Sande}
\affil{Sydney Institute for Astronomy, School of Physics, A28, The University of Sydney, NSW, 2006, Australia}

\author[0000-0001-8823-4845]{David Sobral}
\affil{Department of Physics, Lancaster University, Lancaster LA14YB, UK}
\affil{Leiden Observatory, Leiden Unversity, P. O. Box 9513, NL-2300 AA Leiden, The Netherlands}

\author[0000-0003-3256-5615]{Justin S. Spilker}
\affiliation{Department of Astronomy, University of Texas at Austin, 2515 Speedway, Stop C1400, Austin, TX 78712, USA}

\begin{abstract}

  We present the second data release of the Large Early Galaxy Astrophysics Census (LEGA-C), an ESO 130$-$night public spectroscopic survey conducted with VIMOS on the Very Large Telescope. We release \totn\ spectra with typical continuum $S/N\simeq20\ \rA^{-1}$ of galaxies at $0.6\lesssim z\lesssim 1.0$, each observed for $\sim20$ hours and fully reduced with a custom-built pipeline. We also release a catalog with spectroscopic redshifts, emission line fluxes, Lick/IDS indices, and observed stellar and gas velocity dispersions that are spatially integrated quantities including both rotational motions and genuine dispersion. To illustrate the new parameter space in the intermediate redshift regime probed by LEGA-C we explore relationships between dynamical and stellar population properties. The star-forming galaxies typically have observed stellar velocity dispersions of $\sim150\ \mathrm{km\ s^{-1}}$ and strong \hd\ absorption (\hda$\sim 5\ \rA$), 
  while passive galaxies have higher observed stellar velocity dispersions ($\sim200\ \mathrm{km\ s^{-1}}$) and weak \hd\ absortion (\hda$\sim0\ \rA$). Strong [O{\sc III}]5007$/$\hb\ ratios tend to occur mostly for galaxies with weak \hda\ or galaxies with higher observed velocity dispersion. 
Beyond these broad trends, we find a large diversity of possible combinations of rest-frame colors, absorption line strengths and emission line detections, illustrating the utility of spectroscopic measurements to more accurately understand galaxy evolution. 
  By making the spectra and value-added catalogs publicly available we encourage the community to take advantage of this very substantial investment in telescope time provided by ESO.

\end{abstract}

\keywords{galaxies: evolution --- techniques: spectroscopic --- catalogs --- surveys}

\section{Introduction} \label{sec:intro}

The Sloan Digital Sky Survey \citep[SDSS;][]{York00} is the largest and most widely used legacy dataset of detailed spectroscopic properties of galaxies beyond individual case studies in the present-day Universe. It includes hundreds of thousands of spectra of low redshift galaxies, with high enough signal-to-noise ($S/N$) to allow key measurements of, e.g., the ages and metal content of stellar populations \citep[e.g.][]{Kauffmann03b,Gallazzi05}, structure and size \citep[e.g.][]{Kauffmann03c,vdWel09}, nuclear activity \citep[e.g.][]{Kauffmann03a,Schawinski09}, and internal dynamics \citep[e.g.][]{Gallazzi06,Graves09}. These were further characterized in various scaling relations and diagnostic diagrams, such as (naming only a few examples) the \citet{Baldwin81} diagram \citep[see also][]{Kewley06}, the relations between metallicity, age, and mass \citep{Tremonti04,Gallazzi05}, the relation between luminosity, size and dynamical properties \citep[fundamental plane; e.g.][]{Djorgovski87,Bernardi03}, and various star-formation rate (SFR) scaling relations \citep[e.g.][]{Brinchmann04}, which greatly improved our understanding of galaxy formation and evolution.

{Since then, l}arge multi{-}wavelength photometric surveys have rapidly pushed the information boundary to the distant Universe, so that now the stellar masses, star-formation rates and morphologies of many thousands of galaxies have been measured for mass limited samples up to $z\simeq4$ \citep[e.g.,][]{Grogin11,Koekemoer11,Muzzin13a,Skelton14,Straatman16}. Spectroscopic surveys face a trade-off between sample size and $S/N$. Surveys such as DEEP2 \citep{Newman13}, zCOSMOS \citep{Lilly07}, VIPERS \citep{Guzzo14}, VVDS \citep{leFevre05}, MOSDEF \citep{Kriek15} and KBSS \citep{Steidel14} have already targeted tens of thousands of galaxies up to $z\sim3$ with $8-10-$meter class telescopes, obtaining redshifts and information on intrinsic galaxy properties mainly through bright emission lines originating from ionized gas. Knowledge of more detailed physical properties, especially those that are manifested by continuum emission and absorption lines, such as ages, stellar kinematics, and stellar metallicites, have as of yet been obtained at $z<1$ {either} from relatively small samples ($\lesssim100$) optimized for data quality \citep[e.g.,][]{Kelson01,Treu05,vdWel05,vDokkum07,Moran07,Jorgensen13,Gallazzi14,Bezanson15a}, or from stacks \citep[e.g.,][]{SanchezBlazquez09,Choi14,Siudek17}, {or published as sample averages \citep{Moresco10}}. Beyond $z>1$ observations are limited to stacks \citep[e.g.,][]{Whitaker13,Onodera15}, (lensed) single galaxies \citep[e.g.,][]{vDokkum10,Newman15,Hill16,Kriek16,Toft17,Newman18} or small samples of only a handful to a few tens of sources \citep[e.g.,][]{Newman10,Toft12,vdSande13,Bezanson13,Belli14,Belli15,Belli17}. Therefore the full range of physical properties of galaxies that dominate the stellar mass budget at $z\sim1$ has not yet been studied for individual galaxies in samples sufficiently large to be even remotely comparable with SDSS.

The Large Early Galaxy Astrophysics Census (LEGA-C) survey has been designed to provide both the depth and sample size needed to study at $z=0.6-1$ the scaling relations that have been characterized so well by SDSS. From December 2014 to March 2018, 20 hour deep integrations have been obtained for over 4000 targets with the Visible Multi-Object Spectrograph \citep[VIMOS;][]{leFevre03} on the eight meter Very Large Telescope at Paranal in Chile.

The survey strategy and science goals have been presented by \citet{vdWel16} (hereafter: W16), accompanied by a first release of 892 spectra and corresponding spectroscopic redshifts. In this paper we present Data Release II{\footnote{\url{http://www.eso.org/sci/publications/announcements/sciann17120.html}}}\textsuperscript{,}\footnote{\url{http://www.mpia.de/home/legac/}} (hereafter: DR2). 
This release consists of fully reduced spectra in one dimension of \totn\ sources as well as a catalog with spectroscopic redshifts, emission line fluxes, absorption indices, and (spatially integrated) observed velocity dispersions. {Based on the quantities in this paper's catalog, \citet{Wu18} presented the relation between \hda\ (see Section ~\ref{sec:science}) and \dfn\ \citep{Kauffmann03b}.}

The structure of the paper is as follows: in Sections \ref{sec:sample} and \ref{sec:observations} we discuss the sample selection and observation strategy of LEGA-C. In Section~\ref{sec:pipe} we describe the data reduction process. The contents of DR2 are presented in Section~\ref{sec:release}. Finally, in Section~\ref{sec:science} we present relations between observed velocity dispersion and various star-formation activity and age indicators. Throughout, we use in-air wavelengths and we assume a standard $\Lambda$CDM cosmology with $\Omega_{\mathrm{m}}=0.3$, $\Omega_{\Lambda}=0.7$ and $H_0=70\ \mathrm{km\ s^{-1}\ Mpc^{-1}}$. The adopted photometric system is AB \citep{Oke95}.

\section{Data}
\subsection{Sample selection and completeness}\label{sec:sample}

\begin{figure*}
  \begin{center}
    \includegraphics[width=0.8\textwidth]{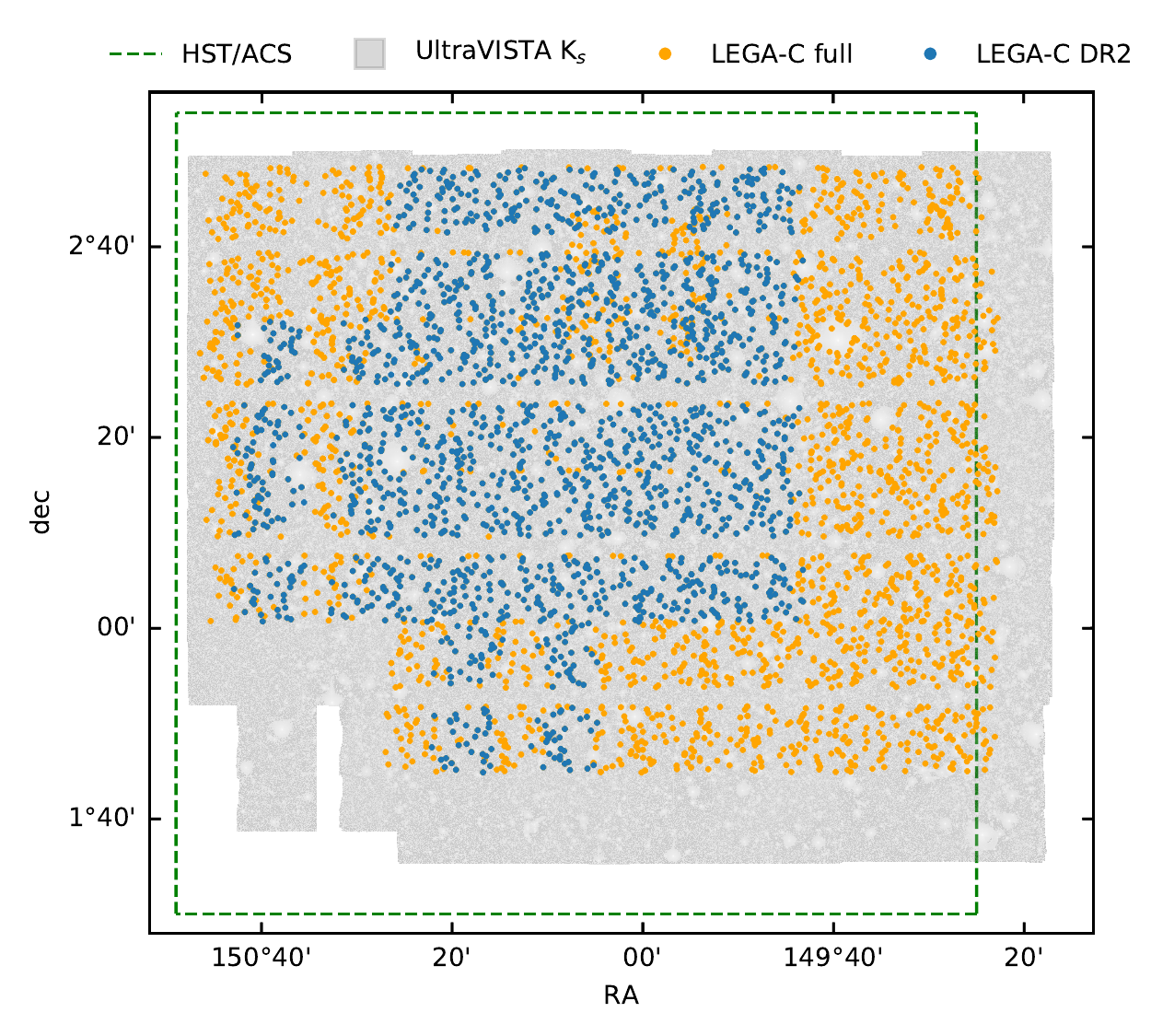}
    \caption{Footprint of the Large Early Galaxy Astrophysics Census survey. Sources were $K_s-$band magnitude selected from the UltraVISTA survey (image in background). Those already included (\totn) in DR2 are shown in blue, while those of the final release ($\gtrsim4000$) are shown in orange. The sources lie in the well-known COSMOS field, the outline of which (as observed with the HST ACS/WFC) is indicated by a dashed frame. None of the LEGA-C masks were chosen to cover regions with particular environmental properties.}
    \label{fig:footprint}
  \end{center}
\end{figure*}

\begin{figure}
  \begin{center}
  \includegraphics[width=0.49\textwidth]{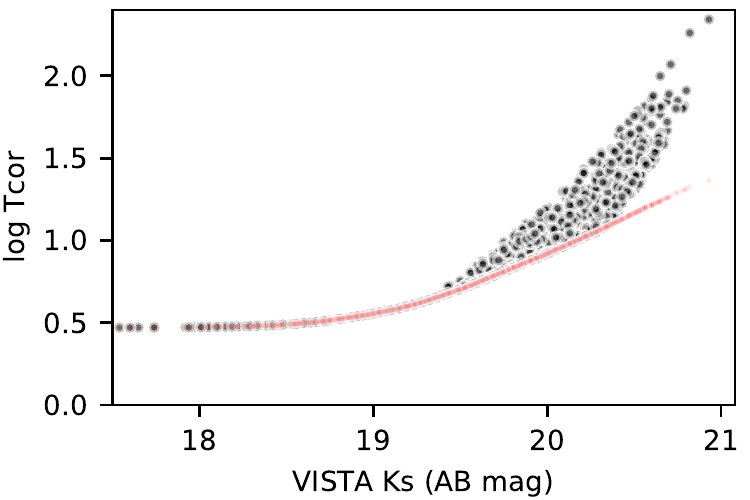}
  \caption{The total completeness correction, Tcor (black dots), provides a way of weighting results for studies based on LEGA-C. It not only accounts for the probability that a galaxy was selected as a target (indicated by red dots), as a function of apparent $K_s-$band magnitude, but also includes a standard cosmological volume correction. The total correction has been included in the catalog for the \prims\ primary targets.}
  \label{fig:1}
  \end{center}
\end{figure}

The primary targets are chosen from a $K_s-$band magnitude selected parent sample of $\sim10\ 000$ galaxies with photometric redshifts $z=0.6-1$ drawn from the Ultra Deep Survey with the VISTA telescope (UltraVISTA) catalog \citep{Muzzin13a}, which overlaps for the most part with the already extensively photometrically covered COSMOS field (Figure~\ref{fig:footprint}). The $K_s-$band limit ranges from $K<21.08$ at $z=0.6$ to $K<20.36$ at $z=1.0$. Overall this results in a sample with a stellar mass limit of the order of  $\sim10^{10}\ \msun$. Targets for the VIMOS masks were inserted in order of $K_s-$band magnitude (brightest first) while avoiding slit collisions. This results in a sample dominated by galaxies with stellar masses $10.5<\logm<11$ and observed stellar velocity dispersions $100<\sigma'<200$ $\mathrm{km\ s^{-1}}$. Among galaxies in the parent sample the probability of inclusion in the survey only depends on $K_s-$band magnitude, such that the sample completeness correction is simply the inverse of the fraction of galaxies with a certain $K_s-$band magnitude observed in UltraVISTA. This probability is a smooth function of $K_s-$band magnitude, as can be seen by the curve delineated by the red data points in Figure~\ref{fig:1} and therefore easily parameterized. The minimum completeness correction (maximum probability) is a factor of $\sim3$ ($\sim1/3$), reflecting the fact that so far we have covered about $1/3$ of the full $1.62\ \mathrm{deg^2}$ UltraVISTA parent sample. The final dataset will cover $\sim2/3$. To translate the primary targets into a volume limited sample we also calculated the maximum redshift at which a galaxy would still be included in the parent sample and used this to derive the standard $V_{max}$ correction. This $V_{max}$ correction affects the fainter galaxies and causes the tail of galaxies with large total completeness corrections at the faint end in Figure~\ref{fig:1}. Any remaining space in the masks was used to include `fillers'. Ordered by priority these fillers are 1) galaxies at $z>1$ and brighter than $K=20.4$; 2) galaxies at $z<1$ and fainter than our K-band magnitude limit. Both of these subsamples are also ranked by $K_s-$band magnitude when designing the masks. The sample to be released in DR2 consists of \prims\ primary targets and \fills\ fillers. Together these are shown in Figure~\ref{fig:footprint} as blue sources. The full LEGA-C sample will comprise $\gtrsim4000$ 
sources (shown as orange sources in Figure~\ref{fig:footprint}). 

\subsection{Observations}\label{sec:observations}

\begin{deluxetable}{rrrrr}
\tablecaption{Masks in LEGA-C Data Release II \label{tab:obs}}
\tablecolumns{5}
\tablenum{1}
\tablewidth{0pt}
\tablehead{\colhead{} & \colhead{} & \colhead{} & \colhead{total integration} & \colhead{seeing}\tablenotemark{a} \\[-6pt]
  \colhead{mask} & \colhead{\# targets} & \colhead{\# primaries} & \colhead{time (hours)} & \colhead{(arcsec)}}
\startdata
1 & 135 & 108 & 21.3 & 1.18 \\ [-3pt] 
2 & 129 & 104 & 19.7 & 0.67 \\ [-3pt] 
3 & 132 & 103 & 21.7 & 0.92 \\ [-3pt] 
4 & 131 & 108 & 21.3 & 1.02 \\ [-3pt] 
5 & 134 & 121 & 21.4 & 1.07 \\ [-3pt] 
6 & 129 & 106 & 21.5 & 1.02 \\ [-3pt] 
7 & 135 & 107 & 22.5 & 0.93 \\ [-3pt] 
8 & 134 & 100 & 21.1 & 1.04 \\ [-3pt] 
9 & 134 & 99 & 20.0 & 0.83 \\ [-3pt] 
10 & 135 & 97 & 20.0 & 1.01 \\ [-3pt] 
11 & 134 & 112 & 21.8 & 0.65 \\ [-3pt] 
12 & 134 & 95 & 20.2 & 0.91 \\ [-3pt] 
13 & 135 & 102 & 26.9 & 0.89 \\ [-3pt] 
14 & 124 & 91 & 19.0 & 0.95 \\ [-3pt] 
15 & 133 & 97 & 23.8 & 0.69 \\  
\hline 
all & \totn & \prims & 322.1 & \\ 
\enddata
\tablenotetext{a}{Taken as the median of the conditions observed at the start of each OB.}
\end{deluxetable}

In this release we present the data from masks 1 through 15 out of the total 32 masks. Details on the mask design can be found in W16. The masks were observed during the ESO observing periods 94 to 98, from December 2014 to January 2017. Due to visibility constraints, most of these nights were fractional allocations to LEGA-C. Observations were conducted in visitor mode, during dark time periods, and in good weather conditions (seeing $\lesssim1.3\ \arcsec$).

Every mask was observed during 15-35 Observing Blocks (OBs) of four $450-900$ second long exposures, depending on observing conditions. A mask was completed after $\sim 20$ hours depending on the conditions. We find minimal fringing, therefore no dithering was applied to avoid severe efficiency costs. 
Dithering would sacrifice $16\%-30\%$ in depth, equivalent to $6-10$ hours of exposure time, as outlined in W16. A precise overview of the number of targets in each mask and the corresponding total integration time and average seeing is given in Table~\ref{tab:obs}.

Calibrations were obtained in a non-standard manner, again to maximize survey efficiency. Initially, arcs and screen flats were not taken directly after each science OB, but at the end of the (partial) night and at the average of the rotator angles of each OB observed that night. However, subsequent analysis of the full calibration dataset showed that proximity in time to the science exposures is more relevant to minimize the effects of flexure and hysteresis than the rotator angle, so our strategy was changed to taking calibration exposures every two to two and a half hours between science OBs. For short partial nights, i.e., less than two hours of allocated time, we only obtained calibration data once before or after observing the science OBs.

We do not use spectrophotometric standards to flux-calibrate the spectra, but rather use the rich and well-calibrated photometric spectral energy distributions (SEDs) from the UltraVISTA catalog to derive the calibration (see Section~\ref{sec:pipe2}).

\subsection{Data reduction}\label{sec:pipe}

\subsubsection{ESO pipeline}

The raw frames were obtained from the ESO archive and processed through the recommended ESO tools at the time of observation: Reflex versions 2.7-2.8.5 \citep{Freudling13} with the VIMOS pipeline package versions 2.9.15-3.1.9. The pipeline, in short, first median combines the 5 raw bias frames corresponding to an observation into a master bias frame after removing their overscans. The master bias is then subtracted from 5 raw flat field frames, the corresponding arc lamp exposure taken at roughly the same time as the flat fields, and the raw science frames. The flat field frames are used to trace the edges of the spectra to find any amount of spatial curvature caused by the CCD, which is then fit with a low-degree polynomial to rectify the arc lamp exposure and science frames. The science frames are divided by a master flat field, a normalized combination of the 5 individual flat fields. 2D spectra are extracted for each individual slit using the extraction window derived from the flat fields and subsequently remapped and wavelength calibrated using an optical distortion solution and wavelength solution both obtained from the arc lamp exposure. 
Spectra are cleaned of cosmic rays after identifying groups of pixels with excess flux compared to the background noise. 
Sky subtraction is deliberately turned off to be addressed in detail in a custom pipeline (see next Section). For each OB, the individual (usually 4) subexposures are averaged, at which point they are ready to be used as input for the custom pipeline.

\subsubsection{Custom LEGA-C pipeline}\label{sec:pipe2}
We employ a custom pipeline to further analyze the Reflex data products. This is motivated by the need to remove bleeding of bright skylines between slits, correct for small misalignments between science and calibration exposures due to instrument flexure, the need for accurate sky-subtraction for extended objects, and optimal $S/N-$weighted extraction of spectra from the different OBs. An initial version of this custom pipeline was used for Data Release I and described in W16. For Data Release II a number of refinements are added and all spectra, including those from Data Release I, were processed or re-processed using the latest version of the pipeline. The custom pipeline performs the following steps:
\begin{enumerate}
  \item Variance spectra are calculated from the (background $+$ object) flux and the read noise on a pixel by pixel basis.
  \item Slit definitions, i.e., the locations of sources, are verified and if necessary adjusted, with a margin at the top and bottom of the slit against bleeding of skyline flux from neighbouring slits.
  \item The location and spatial extent (FWHM) of the sources are measured by fitting a Gaussian in the spatial direction after summing along $\sim800$ pixels in the wavelength region between $7120\rA$ and $7600\rA$, which is free of bright atmospheric emission lines. The location of the source is then traced along the entire wavelength range in bins of 100 pixels through a weighted fit using the variance spectrum. If there is insufficient flux to do this{ (9\% of all individual OBs)}, the location over the entire wavelength range is assumed to be the expected location from the mask design.
  \item We then build a galaxy$+$sky model for each wavelength pixel. This includes the Gaussian with a fixed location and FWHM as defined above, but with the amplitude as a free parameter, and with a constant (sky) flux level as a second free parameter. To account for areas with high sky levels or large sky level gradients in the wavelength direction, e.g., near the edges of skylines, a first-order term is added to allow for a sloping background in the spatial direction. 
  \item Telluric absorption features are corrected by dividing by a normalized blue star spectrum. Each mask includes a blue star, selected to have an SDSS color $(g-r)<0.5$, corresponding to spectral type F or earlier. We only applied corrections in wavelength regions with significant telluric absorption ($6865-6960\rA$, $7160-7330\rA$, $7580-7710\rA$, $8070-8400\rA$, $8900-9200\rA$), so that stellar features outside those windows do not lead to spurious features in the galaxy spectra. 
  \item Spectra from individual OBs are aligned and co-added after weighting by the $S/N$ as measured between $7120\rA$ and $7600\rA$. {We note that for the majority (76\%) of sources for which a location prior had to be used for at least 75\% of the individual OBs due to lack of SNR (step 3), a continuum was detected after co-adding.}
  \item Many galaxies have faint, extended wings in our deep spectra that are not accurately captured with the Gaussian extraction kernel that we use to model the sky$+$object in the individual OBs. We therefore revisit the sky subtraction in the coadded 2D spectra, which have a much higher S/N than the single OB spectra and allow for a better assessment of the extended light. Summing over all wavelength pixels between $7120\rA$ and $7600\rA$ we fit a Moffat profile to the flux of the source. Keeping the shape of the Moffat profile fixed we then create a new model for sky$+$object for each column of pixels with the same wavelength, where we vary only the amplitude and sky background (constant$+$slope). Third-order polynomials are then fit to the sky parameters to create a smooth master sky model which is then subtracted from the 2D spectrum and propagated into the 1D extracted spectrum by using the Moffat profile as the extraction kernel.
  \item Secondary objects are identified automatically in the coadded spectra by searching for significant peaks of continuum flux. If a spectrum contains a secondary source, the sky background in the previous steps will have been overestimated, so we perform a second sky subtraction after fitting two Moffat components to the coadded spectrum. 
  \item Emission lines trace the kinematics and spatial structure of gas, which are often different from that of the stellar continuum. For these reasons the Moffat profile that is used to characterize the spatial structure of the continuum light of a galaxy does not accurately reflect the light distribution of emission lines. As such, the sky subtraction near emission lines can be inaccurate. We identify emission lines by searching for significant peaks in smoothed 1D spectra and repeat the sky$+$object Moffat modeling while leaving the spatial location as an additional free parameter across a 40-pixel region around the central wavelength of the emission line.
  \item Spectra are flux calibrated using the photometric SEDs from UltraVISTA. Best-fit spectral templates for these SEDs were obtained with FAST {\citep{Kriek09}} after updating with the spectroscopic redshifts and fit with a 5th order polynomial across the wavelength range of the LEGA-C spectra. Another 5th order polynomial is fit to the LEGA-C spectra and the ratio between the two is used to scale the LEGA-C spectra in a wavelength dependent way. We note that the flux calibration compensates for slit losses assuming that the spectrum is the same in shape across the galaxy, but does not account for wavelength-dependent spatial gradients along the slit. We also note that in this way emission line fluxes receive the same correction, implicitly assuming slit losses are the same for the continuum and the lines.
\end{enumerate}

\section{Release content}\label{sec:release}

\begin{figure*}
  \includegraphics[width=0.49\textwidth]{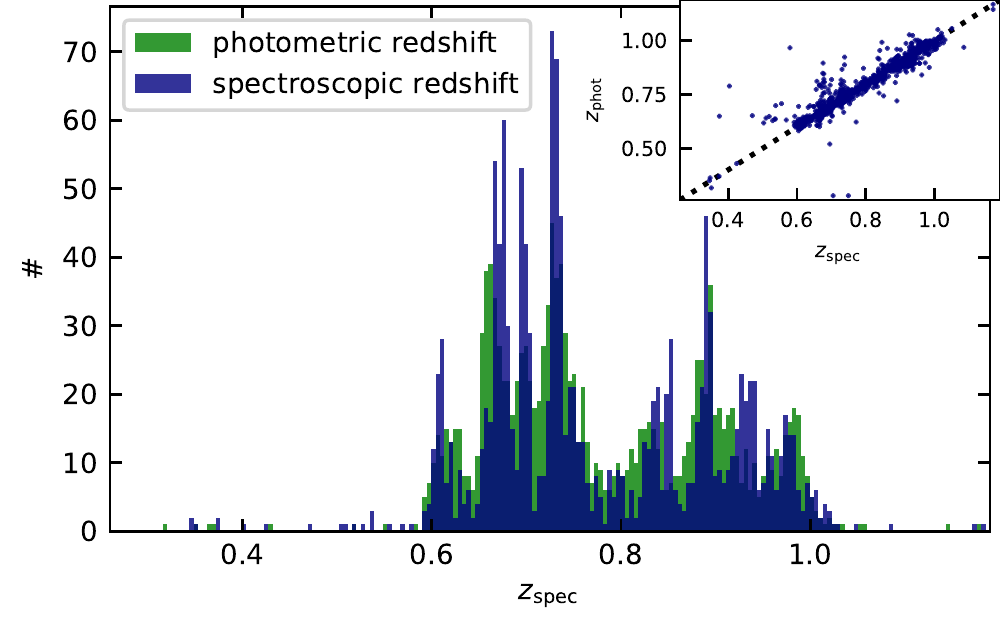}
  \includegraphics[width=0.49\textwidth]{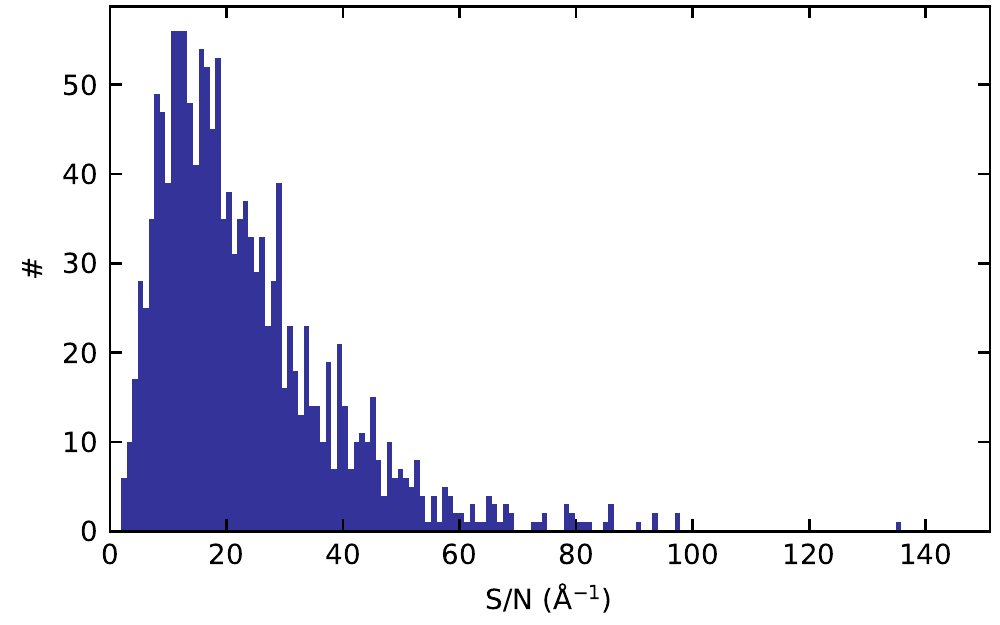}
  \caption{Left: the distribution of spectroscopic redshifts (darkblue histogram) within DR2 of primary targets with $f_{use}=1$ (see Section~\ref{sec:flags}). In the background (green histogram) we show the distribution of photometric redshifts from the UltraVISTA catalog that were used to select the survey targets. The spectroscopic sample is relatively consistent with being drawn from the Ultra VISTA sample, but the spectroscopic redshifts more accurately reveal narrow spikes in the redhift distribution. Right: the $S/N$ distribution, with a median {$S/N=19\ \rA^{-1}$.}}
  \label{fig:hists}
\end{figure*}

The spatially extracted 1D spectra and variance arrays of \totn\ sources with median continuum $S/N=19\ \rA^{-1}$ are released as \texttt{.fits} files, indexed by their respective mask numbers and UltraVISTA IDs. The spectra are accompanied by a catalog, the contents of which will be described in the following subsections. The set of \totn\ spectra is comprised of \prims\ primary targets and \fills\ secondary sources.  
The redshift and $S/N$ distributions are shown in Figure~\ref{fig:hists}. {In general the redshifts agree well (see inset diagram) with a normalized median absolute deviation in $|z_{\mathrm{spec}}-z_{\mathrm{phot}}|/(1+z_{\mathrm{spec}})$ of 0.008 and with only a few strong outliers: $0.8\%$ with $|z_{\mathrm{spec}}-z_{\mathrm{phot}}|/(1+z_{\mathrm{spec}})>0.1$. The outliers tend to have very red rest-frame $V-J$ colors (median $V-J=1.9$) and are caused by the lack of very red SED templates applied in UltraVISTA (see Section~\ref{sec:uvj}). The instrumental resolution for a slit of width $1\arcsec$ is $\mathrm{R}=2500$ at $7500\rA$, but because the galaxies have peaked light profiles -- the slits are not uniformly illuminated -- the effective spectral resolution of our spectra is much better: $\mathrm{R}\sim3500$. We obtained this number by approximating the light profiles of the sources along the slit, truncated by a 5 pixel box car representing the slit width, with a Gaussian.} 

\subsection{Value-added data products}\label{sec:values}

\startlongtable
\begin{deluxetable}{ll}
\tablecaption{DR2 catalog contents \label{tab:cat}}
\tablecolumns{2}
\tablewidth{1pt}
\tablehead{[-20pt]}
\startdata
OBJECT                              & UltraVISTA identifier\\[-3pt]
SPECT\_ID                           & LEGA-C identifier as a unique \\[-3pt]
& combination of OBJECT and mask\\[-3pt]
RAJ2000                             & right ascension (degree)                 \\[-3pt]
DECJ2000                            & declination (degree)                 \\[-3pt]
z                                   & spectroscopic redshift                  \\[-3pt]
Filename                            & file in DR2 \\[-3pt]
SIGMA\_STARS\_PRIME                      & observed stellar velocity dispersion\tablenotemark{a,b}\\[-3pt]
{\footnotesize SIGMA\_STARS\_PRIME\_err}                 & stellar velocity dispersion uncertainty\tablenotemark{a} \\[-3pt]
SIGMA\_GAS\_PRIME                          & observed gas velocity dispersion\tablenotemark{a,b}\\[-3pt]
{\footnotesize SIGMA\_GAS\_PRIME\_err}                     & gas velocity dispersion uncertainty\tablenotemark{a} \\[-3pt]
LICK\_CN1                           & LICK/IDS CN1\tablenotemark{c}        \\[-3pt]
LICK\_CN1\_err                      & LICK/IDS CN1 uncertainty\tablenotemark{c}       \\[-3pt] 
LICK\_CN2                           & LICK/IDS CN2\tablenotemark{c}        \\[-3pt]
LICK\_CN2\_err                      & LICK/IDS CN2 uncertainty\tablenotemark{c}       \\[-3pt] 
LICK\_CA4227                        & LICK/IDS Ca4227\tablenotemark{c}     \\[-3pt]
LICK\_CA4227\_err                   & LICK/IDS Ca4227 uncertainty\tablenotemark{c}     \\[-3pt]
LICK\_G4300                         & LICK/IDS G4300\tablenotemark{c}      \\[-3pt]
LICK\_G4300\_err                    & LICK/IDS G4300 uncertainty\tablenotemark{c}      \\[-3pt]
LICK\_FE4383                        & LICK/IDS Fe4383\tablenotemark{c}     \\[-3pt]
LICK\_FE4383\_err                   & LICK/IDS Fe4383 uncertainty\tablenotemark{c}     \\[-3pt]
LICK\_CA4455                        & LICK/IDS Ca4455\tablenotemark{c}     \\[-3pt]
LICK\_CA4455\_err                   & LICK/IDS Ca4455 uncertainty\tablenotemark{c}     \\[-3pt]
LICK\_FE4531                        & LICK/IDS Fe4531\tablenotemark{c}     \\[-3pt]
LICK\_FE4531\_err                   & LICK/IDS Fe4531 uncertainty\tablenotemark{c}     \\[-3pt]
LICK\_C4668                         & LICK/IDS C4668\tablenotemark{c}      \\[-3pt]
LICK\_C4668\_err                    & LICK/IDS C4668 uncertainty\tablenotemark{c}      \\[-3pt]
LICK\_HB                            & LICK/IDS HB\tablenotemark{c}         \\[-3pt]
LICK\_HB\_err                       & LICK/IDS HB uncertainty\tablenotemark{c}         \\[-3pt]
LICK\_HD\_A                         & LICK/IDS HD\_A\tablenotemark{c}      \\[-3pt]
LICK\_HD\_A\_err                    & LICK/IDS HD\_A uncertainty\tablenotemark{c}      \\[-3pt]
LICK\_HG\_A                         & LICK/IDS HG\_A\tablenotemark{c}      \\[-3pt]
LICK\_HG\_A\_err                    & LICK/IDS HG\_A uncertainty\tablenotemark{c}      \\[-3pt]
LICK\_HD\_F                         & LICK/IDS HD\_F\tablenotemark{c}      \\[-3pt]
LICK\_HD\_F\_err                    & LICK/IDS HD\_F uncertainty\tablenotemark{c}      \\[-3pt]
LICK\_HG\_F                         & LICK/IDS HG\_F\tablenotemark{c}      \\[-3pt]
LICK\_HG\_F\_err                    & LICK/IDS HG\_F uncertainty\tablenotemark{c}      \\[-3pt]
LICK\_D4000\_N                      & LICK/IDS D4000\_N\tablenotemark{c}   \\[-3pt]   
LICK\_D4000\_N\_err                 & LICK/IDS D4000\_N uncertainty\tablenotemark{c}       \\[-3pt]   
Hd\_flux                            & \hd\ emission line flux\tablenotemark{d}             \\[-3pt]
Hd\_err                             & \hd\ line flux uncertainty\tablenotemark{d}\\[-3pt]
Hd\_EW                              & \hd\ equivalent width\tablenotemark{c}                                      \\[-3pt]
Hd\_EW\_err                         & \hd\ EW uncertainty\tablenotemark{c}                         \\[-3pt]
Hg\_flux                            & \hg\ emission line flux\tablenotemark{d}           \\[-3pt]
Hg\_err                             & \hg\ line flux uncertainty\tablenotemark{d} \\[-3pt]
Hg\_EW                              & \hg\ equivalent width\tablenotemark{c}                                     \\[-3pt]
Hg\_EW\_err                         & \hg\ EW uncertainty\tablenotemark{c}                          \\[-3pt]
Hb\_flux                            & \hb\ emission line flux\tablenotemark{d}           \\[-3pt]
Hb\_err                             & \hb\ line flux uncertainty\tablenotemark{d} \\[-3pt]
Hb\_EW                              & \hb\ equivalent width\tablenotemark{c}                                    \\[-3pt]
Hb\_EW\_err                         & \hb\ EW uncertainty\tablenotemark{c}                     \\[-3pt]
OII\_3727\_flux                     & [O{\sc II}] emission line flux\tablenotemark{d}           \\[-3pt]
OII\_3727\_err                      & [O{\sc II}] line flux uncertainty\tablenotemark{d} \\[-3pt]
OII\_3727\_EW                       & [O{\sc II}] equivalent width\tablenotemark{c}                            \\[-3pt]
OII\_3727\_EW\_err                  & [O{\sc II}] EW uncertainty\tablenotemark{c}                 \\[-3pt]
OIII\_4959\_flux                    & [O{\sc III}]4959 emission line flux\tablenotemark{d}           \\[-3pt]
OIII\_4959\_err                     & [O{\sc III}]4959 line flux uncertainty\tablenotemark{d} \\[-3pt]
OIII\_4959\_EW                      & [O{\sc III}]4959 equivalent width\tablenotemark{c}                              \\[-3pt]
OIII\_4959\_EW\_err                 & [O{\sc III}]4959 EW uncertainty\tablenotemark{c}                  \\[-3pt]
OIII\_5007\_flux                    & [O{\sc III}]5007 emission line flux\tablenotemark{d}           \\[-3pt]
OIII\_5007\_err                     & [O{\sc III}]5007 line flux uncertainty\tablenotemark{d} \\[-3pt]
OIII\_5007\_EW                      & [O{\sc III}]5007 equivalent width\tablenotemark{c}                        \\[-3pt]
OIII\_5007\_EW\_err                 & [O{\sc III}]5007 EW uncertainty\tablenotemark{c}          \\[-3pt]
f\_ppxf                             & pPXF fit quality flag (0: good fit)\\[-3pt]
f\_z                                & redshift flag (0: redshift detected)\\[-3pt]
f\_spec                             & spectral quality flag (0: good spectrum)\\[-3pt]
f\_primary                          & primary source flag (1: source primary)\\[-3pt]
f\_use                              & general quality flag (1: good to use)\\[-3pt]
f\_int                              & interpretation flag (0: no issues)\\[-3pt]
Tcor                                & total completeness correction \\ [-3pt]
SN                                  & overall median $S/N$\tablenotemark{e} \\ [-3pt]
SN\_RF\_4000                        & $S/N$ at rest-frame $4000\rA$\tablenotemark{e} \\ [-3pt]
SN\_OBS\_8030                       & $S/N$ at observed frame $8030\rA$\tablenotemark{e} \\ 
\enddata
\tablenotetext{a}{($\mathrm{km\ s^{-1}}$)}
\tablenotetext{b}{measured from the broadening of a spectral line in 1D, this includes\\ both rotation and genuine dispersion}
\tablenotetext{c}{($\rA$)}
\tablenotetext{d}{($10^{-19}\ \mathrm{erg\ cm^{-2}\ s^{-1}}$)}
\tablenotetext{e}{($\mathrm{pix^{-1}}$)}
\end{deluxetable}

DR2 contains spectroscopic redshifts, emission line fluxes and equivalent widths (EWs), Lick/IDS absorption indices, and observed stellar and gas velocity dispersions (see Table \ref{tab:cat}).

The observed stellar and gas velocity dispersions were derived after fitting each spectrum with two template sets using the Penalized Pixel-Fitting (pPXF) code \citep{Cappellari04} using the updated Python routines \citep{Cappellari17}. These consisted 1) of a collection of high resolution ($\mathrm{R}=10\ 000$) single stellar population templates (Conroy et al., in prep.) {downgraded to match the resolution of the LEGA-C spectra} to fit the continuum, as well as  2) a collection of possible emission lines ([Ne{\sc V}], [Ne{\sc VI}], H10, H9, H8, \hd, \hg, \hb, [O{\sc II}]3726,3729, [Ne{\sc III}], [S{\sc II}]6717,6731, [O{\sc III}]5007, [O{\sc III}]4959, [O{\sc III}]4363, [N{\sc I}]) to fit the ionized gas emission. The fluxes were allowed to vary independently, but the linewidths were constrained to be the same \citep[see also W16,][Bezanson et al. in prep]{Bezanson18}. The stellar and gas dispersions are the widths of the absorption and emission lines, respectively. Within a spatially collapsed 1D spectrum the information on rotational motions and genuine intrinsic dispersion is mixed. To emphasize that this is a spatially integrated quantity and to avoid any confusion with intrinsic dispersion we use the term ``observed velocity dispersion'' or the designation $\sigma'$ (as opposed to $\sigma$) throughout this work.   

The emission line fluxes in the catalog were derived from integrating the best-fit emission line set at locations where an emission line was found. EWs were derived by dividing the line flux by the local best-fit continuum, here defined as the average of the continuum in a 200$\rA$ wide region centered on the line. Lick/IDS indices as described in \citet{Worthey94,Worthey97}, including the higher order Balmer lines, \hg$_{\mathrm{A}}$ and \hd$_{\mathrm{A}}$, and the narrow definition of the $4000\rA$-break (\dfn) \citep{Balogh99}, were derived following the analysis in \citet{Gallazzi14} after subtracting the best-fit emission line model. Missing values (due to, e.g., no coverage) are set to NaN in the catalog. For the Lick/IDS indices, {for which there is} sufficient coverage, but $<80\%$ good pixels in the central or pseudo-continuum bandpasses, the values are still included in the catalog, {except} the uncertainty is set to NaN.

\subsection{Uncertainties}\label{sec:unc}

\begin{deluxetable}{rrr}
\tablecaption{Factors applied to formal uncertainties \label{tab:unc}}
\tablecolumns{3}
\tablewidth{0pt}
\tablehead{\colhead{}  & \colhead{correction star-forming} & \colhead{correction quiescent}}
\startdata
SIGMA\_STARS\_PRIME\_err                  & 1.7 & 1.7 \\[-3pt]
SIGMA\_GAS\_PRIME\_err                  & 1.7 & 1.7 \\[-3pt]
LICK\_CN1\_err                       & 3&2  \\[-3pt]
LICK\_CN2\_err                       & 3&2 \\[-3pt]
LICK\_CA4227\_err                    & 2&2 \\[-3pt]
LICK\_G4300\_err                     & 2.5&2  \\[-3pt]
LICK\_FE4383\_err                    & 2.5&2.5  \\[-3pt]
LICK\_CA4455\_err                    & 2& 2 \\[-3pt]
LICK\_FE4531\_err                    & 2.5& 2.5 \\[-3pt]
LICK\_C4668\_err                     & 3&3 \\[-3pt]
LICK\_HB\_err                        & 3& 2 \\[-3pt]
LICK\_HD\_A\_err                     & 2&1.5  \\[-3pt]
LICK\_HG\_A\_err                     & 2&1.5  \\[-3pt]
LICK\_HD\_F\_err                     & 2&1.5  \\[-3pt]
LICK\_HG\_F\_err                     & 2.5&2  \\[-3pt]
LICK\_D4000\_N\_err                  & 2.5&2  \\[-3pt]  
Hd\_err                              & 1.5&1.5 \\[-3pt]
Hd\_EW\_err                          & 1.5&1.5 \\[-3pt]
Hg\_err                              & 2&2 \\[-3pt]
Hg\_EW\_err                          & 2&2 \\[-3pt]
Hb\_err                              & 2.5&2.5 \\[-3pt]
Hb\_EW\_err                          & 2.5&2.5 \\[-3pt]
OII\_3727\_err                       & 2&2 \\[-3pt]
OII\_3727\_EW\_err                   & 2&2 \\[-3pt]
OIII\_4959\_err                      &2 &2 \\[-3pt]
OIII\_4959\_EW\_err                  & 2&2 \\[-3pt]
OIII\_5007\_err                      & 2&2 \\[-3pt]
OIII\_5007\_EW\_err                  & 2&2 \\
\enddata
\end{deluxetable}

Formal uncertainties on the quantities described above were derived from the variance spectra. In general, we find that those agree well with the noise level seen in the spectra themselves. Yet{,} the 
formal uncertainties do not take into account a number of additional sources of error. First, variations in slit alignment after a{c}quisition {as well as} seeing among the different OBs lead to variations in the captured light. Second, the sky$+$object model is imperfect, i.e., a Gaussian or Moffat profile is not an exact representation of a galaxy's light profile. Third, the derived observed velocity dispersions depend on the wavelength range and the choice of templates used to fit the spectra, although pPXF takes into account some uncertainty in the overal spectral shape. Fourth, spectral index measurements are only formally correct for constant variance spectra within the index wavelength bands, and do not take into account variations of several orders of magnitude that arise due to the presence of bright atmospheric emission lines. Fifth, formal uncertainties do not capture uncertainties associated with emission line modeling and subtraction.

Due to these complexities the true uncertainties in the spectra and derivative data products cannot be formally derived. 
Instead we make use of duplicate observations that we have for 44 galaxies in our primary sample; these occur due to slight overlap between the mask pointings. The variation among the duplicate spectra and the measurements based upon those capture all of the additional uncertainties described above and we can use the variance among the duplicate measurements as an independent derivation of our true measurement uncertainties. 

We find that the pixel-to-pixel variation among duplicate spectra exceeds the expected variation based on the variance spectra by a factor $1.25\pm0.25$. That is, the true random uncertainty level is typically 25\% higher than what the variance spectra suggest, but with substantial variation. These reflect the first two sources of uncertainty mentioned above (variations in seeing / alignment and the sky$+$object model imperfections).

As a result we can expect a larger variation in the measurement of, for example, observed stellar velocity dispersions among the duplicates than what the formal uncertainties would suggest. Indeed, we find that the standard deviation among the velocity dispersion measurements is $1.7\times$ larger than expected, which indicates that our velocity dispersion measurements do not only suffer from variations among the duplicate spectra but that the uncertainties are in fact dominated by template mismatch and differences in wavelength coverage.

We analyzed the emission line and absorption index measurements in a similar manner and applied correction factors to the uncertainties that range from 1.5 to 3 (see Table~\ref{tab:unc}). These were sometimes different for quiescent and star-forming galaxies, but mainly depend on the observed wavelength: features in the blue typically have smaller correction factors than features at long wavelengths. 

While these correction factors may seem large it should be kept in mind that the formal uncertainties on, for example, the \dfn\ break are vanishingly small: it is based on averaging hundreds of pixels with individual $S/N > 10$, leading to sub-percent level formal uncertainties. It is no surprise that such precision cannot be actually reached with 20-hour long integrations with varying conditions and unavoidable imperfections in the data processing algorithms. We also note that the Balmer absorption line strengths and their uncertainties are only robust if the Balmer emission lines have flux $<10^{-16}\ \mathrm{erg\ cm^{-2}\ s^{-1}}$.

\subsection{Quality flags}\label{sec:flags}

All spectra, as well as the derived redshifts and observed velocity dispersions, were visually inspected and assigned flags if necessary. A small number of spectra were rejected on the basis of fundamental flaws (e.g. imperfect sky subtraction, bad wavelength solution, mismatch between different OBs, absence of light because of vignetting) in the observed spectra or data reduction process and received a flag $f_{\mathrm{spec}} = 1$. For a few galaxies we could not measure the redshift: $f_z =1$, (due to, e.g., vignetting or low $S/N$). We note that in general the spectroscopic redshifts derived from the continuum features are unambiguous for our spectra due to the number of features at high $S/N$, so that there is no need for a quality flag for the redshift measurement as is common practice for redshift surveys. Finally, in a few cases the pPXF fit is flawed., e.g., because absorption features were not detected or the continuum was clearly badly fit: ($f_{\mathrm{pPXF}} =1$). The spectroscopic sample of primary survey targets that can be used for scientific purposes ($f_{use}=1$) combines these flags: $f_{\mathrm{primary}} =1$ and $f_{\mathrm{spec}} =0$ and $f_{\mathrm{z}} =0$ and $f_{\mathrm{pPXF}} =0$. 
\nuse\ of \prims\ primaries have $f_{use}=1$. 

\begin{figure*}
  \begin{center}
    \includegraphics[width=0.8\textwidth]{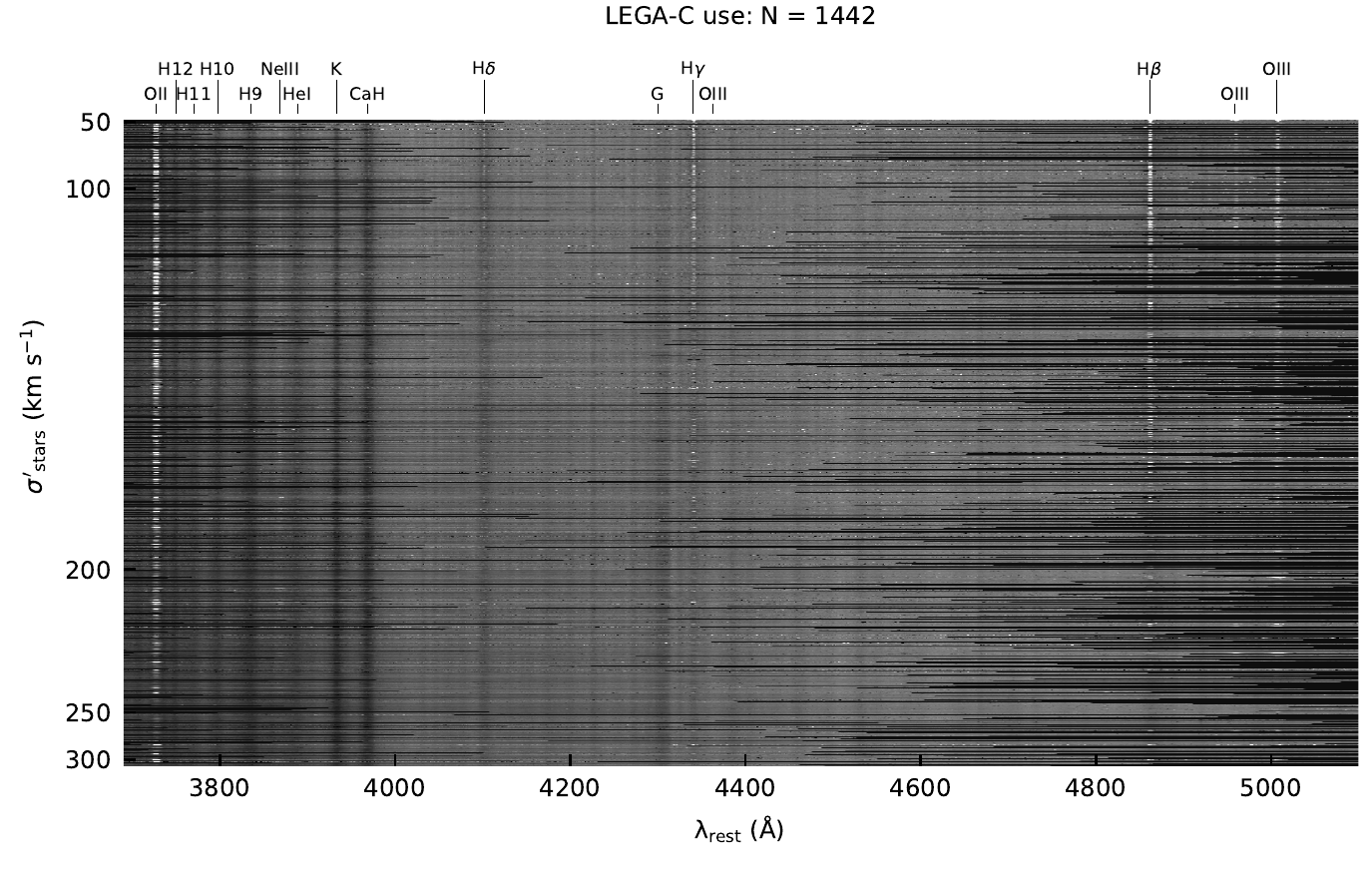}
    \caption{Stack of reduced spectra with $f_{use}=1$. To create the stack, the individual spectra were normalized, smoothed with a 5 pixel wide box filter and resampled to rest-frame wavelengths, and sorted according to their observed stellar velocity dispersion. Zeroes (dark colors) were used for rest-frame wavelengths that were not covered. The many emission and absorption features are clearly visible even if they are overlapping (e.g., \hda\ at low observed velocity dispersion) and can be seen to vary according to physical properties, e.g. Balmer emission line strength decreases towards increasing velocity dispersion, likewise G4300 absorption increases.}
    \label{fig:all}
  \end{center}
\end{figure*}

We show a stack of the released spectra with $f_{use}=1$ in Figure~\ref{fig:all}, sorted by observed stellar velocity dispersion. The stack was created after smoothing with a five pixel wide box filter and resampling each spectrum to a common grid ranging from $3690\rA$ to $5100\rA$ in rest-frame wavelength space and normalizing by the maximum flux of the corresponding best-fit pPXF model. The Figure clearly shows the wide range and variety of emission and absorption lines as well as the sensitivity of our data to changes in the physical properties of the galaxies. For example, we observe decreasing Balmer emission line strengths towards higher observed velocity dispersions, or, e.g., stronger G4300 absorption. In some cases both the absorption line and emission line infill are clearly visible (e.g., He{\sc I}, CaH, \hd). Examples of individual spectra are shown in Figure~\ref{fig:spectra} in the Appendix.

In addition to spectral quality flags we release an interpretation flag ($f_{int}$), which is set to one if the interpretation of any of the value added quantities needs extra attention. For example, in the case of a merging system, the measured quantities may be derived from the sum of the spectra of two galaxies. Since the quality of the spectra in these cases is good, $f_{int}$ is a separate flag from $f_{use}$ and some sources with $f_{use}=1$ may also have $f_{int}=1$.

\section{Connecting dynamical and stellar population properties}\label{sec:science}

In this section we illustrate the new parameter space that can be explored with the LEGA-C dataset: until now it was not possible to examine at once at higher redshift the dynamical and stellar population properties of large samples of galaxies of all types.

\subsection{Observed stellar velocity dispersion versus \hda}

\begin{figure*}
  \begin{center}
  \includegraphics[width=0.63\textwidth]{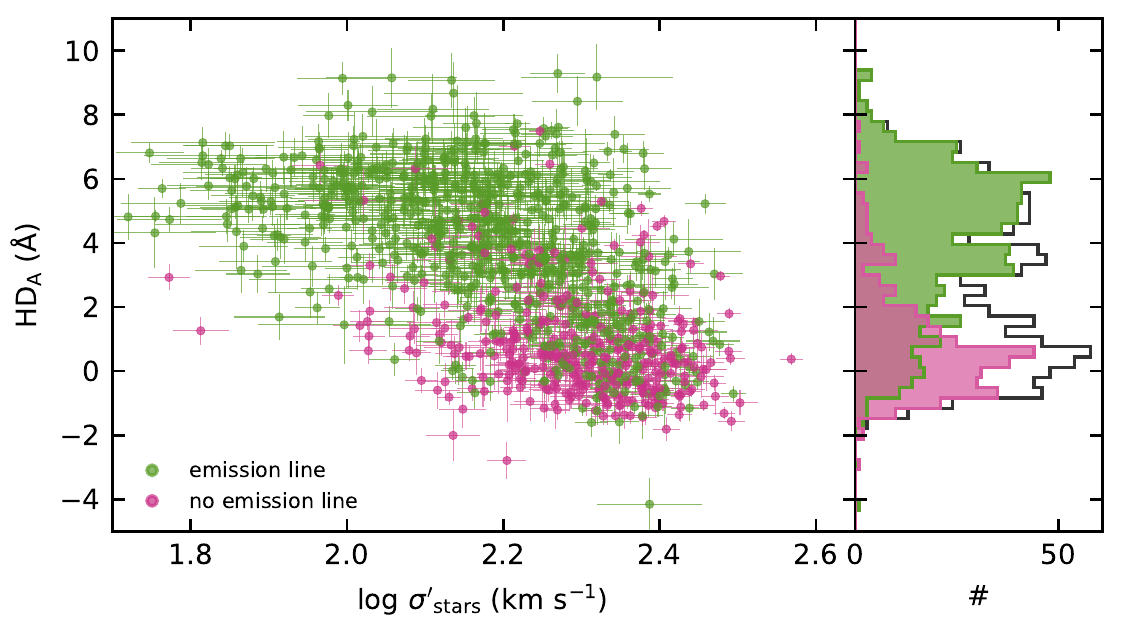}
  \caption{\hda\ versus stellar observed velocity dispersion for {1071} galaxies. Data are color coded green if an emission line (any of those included in the catalog; see Table~\ref{tab:cat}) has been detected at $5\sigma$. Galaxies with higher observed velocity dispersions tend to have lower \hda\ and fewer emission line detecions, {while the distribution of detected and non-detected emission lines traces out a bi-modal distribution in \hda.}}
  \label{fig:min4}
  \end{center}
\end{figure*}

In Figure~\ref{fig:min4} we show observed stellar velocity dispersion (catalog: \texttt{SIGMA\_STARS\_PRIME}) versus \hda\ (catalog: \texttt{LICK\_HD\_A}) for a sample {of 1071 galaxies} selected to have $f_{use}=1$, $f_{int}=0$ and $S/N$ measured at $4000\ \rA$ of $>10\ \mathrm{pix^{-1}}$.
We investigate observed stellar velocity dispersion rather than observed gas velocity dispersion, as the central potential well of a massive galaxy is dominated by stars and intrinsic ionized gas dispersion does not trace collisionless orbits, although Bezanson et al. (in preparation) show that the two quantities are very similar within LEGA-C.  \hda\ is an absorption feature mainly present in the spectra of A$-$type stars. It can be used as an age indicator of the stellar population, although at a fixed age more metal-rich populations may exhibit stronger \hd\ absorption as well. In this Subsection and the next we ignore metallicity effects, but in Section~\ref{sec:agemetal} we will discuss the potential of LEGA-C to investigate age and metallicity effects using the wealth of high SNR absorption features detected in the spectra. 

As can be seen in Figure~\ref{fig:min4}, the galaxies follow a broad trend, with higher observed velocity dispersions for galaxies with lower \hda. Although galaxies at the high dispersion end ($>200$ $\mathrm{km\ s^{-1}}$) can exhibit a range of ages, as traced by \hda, they appear to be predominantly old (\hda$<2\ \rA$). Galaxies with high \hda\ may have large observed velocity dispersions ($>200$ $\mathrm{km\ s^{-1}}$), but those are relatively few. Similarly, there are few old galaxies with observed velocity dispersions $<150$ $\mathrm{km\ s^{-1}}$.

Some galaxies scatter away from the bulk of the sample. There is a small group of apparent outliers with \hda$>8\ \rA$, but these have values within the range predicted by SFH models to be occupied by very young, actively star-forming galaxies \citep[e.g.,][]{Gallazzi14,Wu18}. The galaxy with \hda$\sim-4$ has a photometric SED typical of a very old quiescent galaxy, but the spectrum has relatively low  $S/N(4000\ \rA)=11.2\ \mathrm{pix^{-1}}$ and some weak [O{\sc II}] emission, and \hda\ may have been underestimated if residual \hd\ emission was not captured. 
Galaxies that scatter to the left of the majority of the datapoints tend to have relatively large axis-ratios, indicating that at least some are viewed face-on and we are potentially missing a rotational component contributing to the observed velocity dispersion. Examples of galaxies from the various regions in $\sigma'-$\hda\ parameter space are shown in Figure~\ref{fig:spectra} in the Appendix.

The trend between age (or metallicity) and observed velocity dispersion is also evident from the lack of emission lines detected in galaxies with high observed velocity dispersions and low \hda. The distribution of \hda\ is weakly bi{-}modal, but comparing emission line detections and non-detections we find two distinct distributions. For galaxies without detected emission lines this suggests only very low levels of star-formation. Even though the strongest Balmer line, \ha, which is the bright diagnostic used in most surveys, is not within the wavelength coverage of the LEGA-C primary sample, the median uncertainty on \hb\ flux in LEGA-C is only \hblim, indicating a $1\sigma$ upper limit for non-detections of SFR$=$\sfrlima\ \citep{Kennicutt98} at their median redshift \medianz. For \hg\ and \hd\ emission similar upper limits can be derived of \sfrlimb\ and \sfrlimc, respectively, so we speculate that galaxies undetected in any of these three lines (the chance of skyline contamination of all three at the same time and/or lack of wavelength range is small) have at most an unobscured SFR of \sfrlimfinal\ at the $3\sigma$ level, although a non-detection could also indicate dust obscuration.

Some galaxies have \hda$>2$, but still no detected emission lines. These tend to have red rest-frame $U-V$ colors ($U-V>1.3$ for all but two) and are further described in Section~\ref{sec:uvj}.

\begin{figure}
  \includegraphics[width=0.49\textwidth]{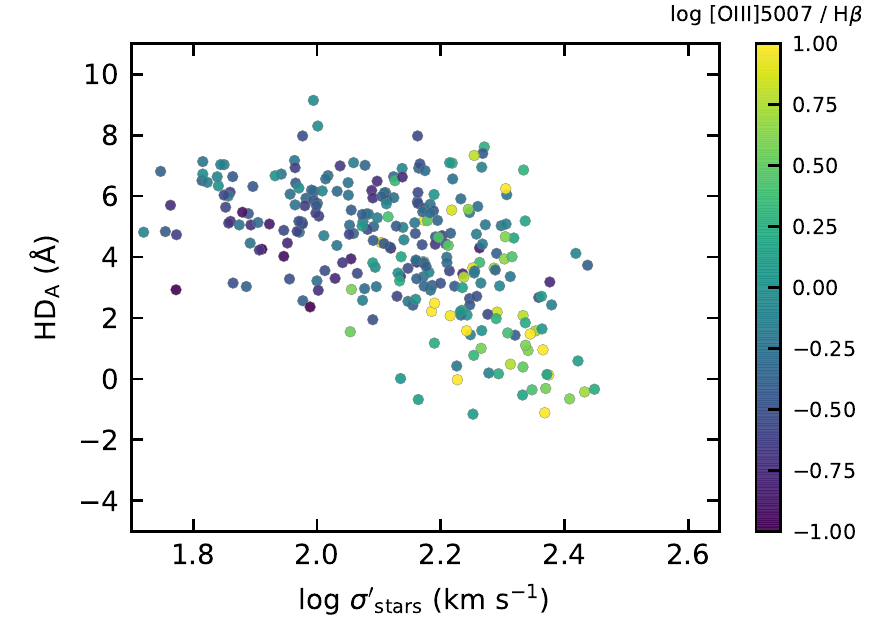}
  \caption{\hda\ versus observed stellar velocity dispersion, color-coded by the line ratio [O{\sc III}]5007$/$\hb. Strong [O{\sc III}]5007$/$\hb\ occurs for older (\hda$<2$) galaxies or younger galaxies with relatively high observed velocity dispersions.}
  \label{fig:min2}
\end{figure}

In Figure~\ref{fig:min2} we show the ratio [O{\sc III}]5007$/$\hb\ (catalog: \texttt{OIII\_5007\_flux} and \texttt{Hb\_flux}) for sources selected depending on the rest-frame wavelength coverage of the spectra. 
We find occurrences of an excess log [O{\sc III}]5007$/$\hb$>0.5$ in a modest number of galaxies with $\sigma'_{\mathrm{stars}}>150$ $\mathrm{km\ s^{-1}}$ and \hda$>2$ and for most galaxies with \hda$<2$. This may suggest that these galaxies more often harbour active galactic nuclei (AGN) or low-ionization nuclear emission-line regions \citep[LINERs, e.g.,][]{Heckman80,Baldwin81,Kewley06}.

\subsection{Characterizing galaxy properties within the UVJ-diagram}\label{sec:uvj}

In this Section we examine the previously shown quantities in the context of the UVJ diagram. Within the UVJ diagram \citep[e.g.,][]{Labbe05,Williams09} galaxies exhibit a bi{-}modal distribution, with quiescent galaxies populating a small region in the top-left of the diagram (red rest-frame $U-V$ and blue $V-J$ colors). As such, a galaxy's rest-frame $U-V$ and $V-J$ color can be used to determine if it is star-forming or quiescent, a method which has been shown to be effective on average for galaxies up to $z\simeq3$ \citep{Straatman16}. {In addition, the UVJ diagram can be used to investigate what properties a galaxy may have, given its U-V and V-J colours, but s}{o far, only photometrically derived properties have been investigated in large samples of individual galaxies, such as sSFRs from UV and IR observations \citep[e.g.,][]{Straatman16}, ages, stellar masses, and sSFRs derived from SED fitting \citep[e.g.,][]{Williams10,Whitaker12,Pacifici16}, or average measurements of \ha\ emission line strength or \dfn\ based on stacked SEDs \citep[e.g.,][]{Kriek11,Forrest16}. With LEGA-C, we can for the first time study detailed galaxy population properties based on {deep continuum }spectroscopy of individual galaxies within a $K_s-$band selected sample.

We calculated rest-frame $U-V$ and $V-J$ colors by fitting template spectra to the UltraVISTA photometric SEDs. 
We used the same set of seven templates as \citet{Muzzin13a}, six of which are derived from the PEGASE models \citep{Fioc99} and the last one a red template from the models of \citet{Maraston05}. We note that this set does not include extra templates representing very dust-obscured young galaxies, that in later surveys \citep[e.g.,][]{Skelton14,Straatman16} helped to prevent overestimated photometric redshifts for such galaxies, but we also note that here we fixed the redshifts to the LEGA-C spectroscopic redshifts. Further details about the fitting procedure can be found in \citet{Brammer08}. We excluded 52 extra sources with suspicious photometry due to potentially strong contamination from neighboring sources.

\begin{figure*}
  \includegraphics[width=0.49\textwidth]{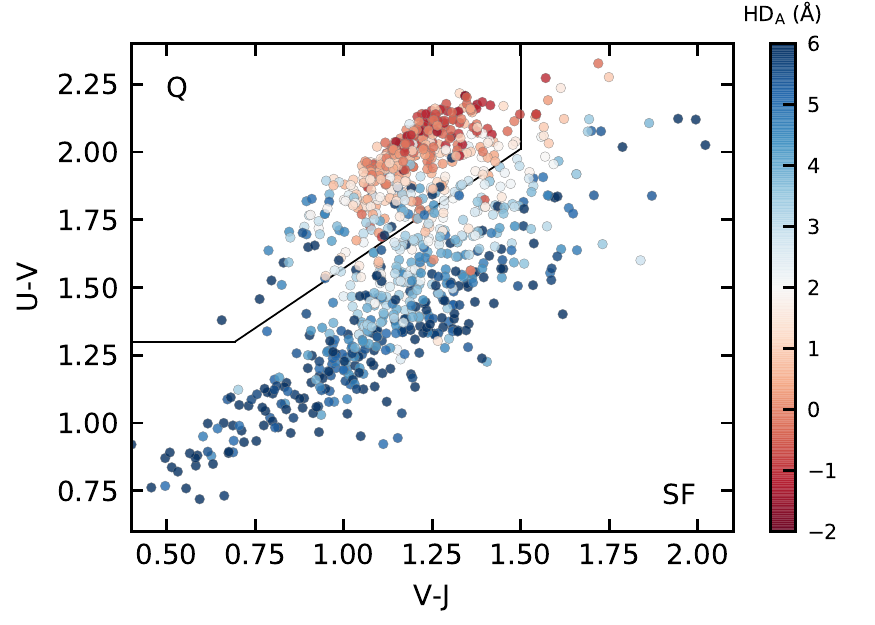}
  \includegraphics[width=0.49\textwidth]{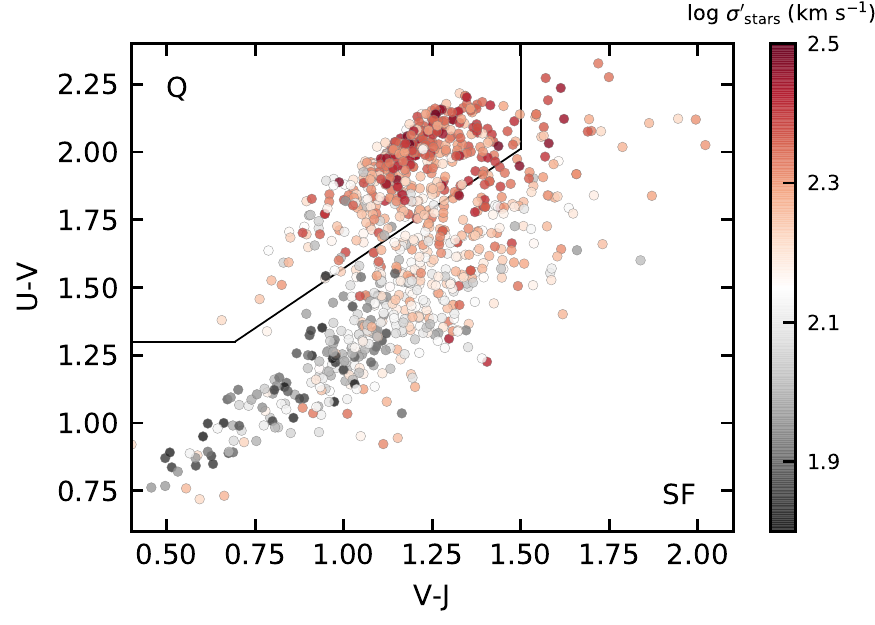}
  \caption{UVJ diagrams color-coded by \hda\ (left-hand panel) or $\sigma'_{\mathrm{stars}}$ (right-hand panel). Quiescent (Q) and star-forming (SF) galaxies are separated using the definition of \citet[][black line]{Muzzin13b}. The age indicator \hda\ decreases towards redder $U-V$ and bluer $V-J$ colors, with the lowest values of \hda\ occuring for quiescent galaxies (the sequence of datapoints in the top-left region of the diagrams). On the red sequence \hda\ decreases again towards redder $U-V$. Observed velocity dispersion follows a different trend compared with \hda, such that $\sigma'_{\mathrm{stars}}$ generally increases towards redder $U-V$. 
    In terms of \hda\ we find a distinct subpopulation of younger quiescent galaxies with bluer $U-V$ and $V-J$ colors. These could be post-starburst galaxies that may only recently have transitioned out of the star-forming population. These galaxies appear to have relatively low $\sigma'_{\mathrm{stars}}$ as well compared to other quiescent galaxies, albeit in a less distict way.}
  \label{fig:min1}
\end{figure*}

In the left panel of Figure~\ref{fig:min1} we examine \hda\ and find that it is strongly related to whether a galaxy is photometrically defined as quiescent or star-forming, with lower values of \hda\ for quiescent galaxies. Within the star-forming population \hda\ can have a range of values, but there appears to be a trend of lower \hda\ towards redder $U-V$ and bluer $V-J$, that is, towards the locus of quiescent galaxies. This is similar to the previously observed trend of decreasing sSFR based on broadband data \citep[e.g.,][]{Williams10,Straatman16,Fang18}, although we note that the value of \hda\ is affected both by age and metallicity. Within the quiescent population, \hda\ decreases with increasing $U-V$, indicating an older population, consistent with similar results from \citet{Whitaker12}. We also find within the quiescent region of the UVJ diagram a distinct population with relatively blue $U-V$ and $V-J$ colors and high \hda\ values comparable to those of star-forming galaxies. {We identify 16 galaxies with $3<$\hda$(\rA)<6$ (indicative of ages of $\sim1$ Gyr; see also Section \ref{sec:agemetal}) and no emission lines (see Figure \ref{fig:uvjem} below).} They likely have much younger ages than the bulk of the quiescent population, which in turn suggests they have only recently, but rapidly, transitioned from star-forming to quiescent. These galaxies may be similar to those often designated as ``post-starburst'', with strong Balmer absorption features resembling spectra of A$-$type stars, indicating a recent burst of star-formation $\sim1$ Gyr earlier and a subsequent sudden cessation of star-formation \citep[e.g.,][]{Quintero04,Goto05,LeBorgne06,Wild09,Whitaker12,McIntosh14}.

In the right panel of Figure~\ref{fig:min1} we examine observed velocity dispersion. 
Along the locus of the star-forming sequence observed velocity dispersion increases towards redder $U-V$ and redder $V-J$. This is the same trend as observed for stellar mass \citep[again, see, e.g.,][]{Williams10,Straatman16,Fang18}. The lowest observed velocity dispersions are found for the bluest galaxies. Quiescent galaxies in the UVJ diagram have relatively high observed velocity dispersions. These increase with increasing $U-V$, with relatively low $\sigma'_{\mathrm{stars}}$ for the post-starburst population. We also find that the highest dispersions occur not only for quiescent galaxies, but also for red (in $V-J$ and $U-V$) star-forming galaxies. 

\begin{figure*}
  \includegraphics[width=0.49\textwidth]{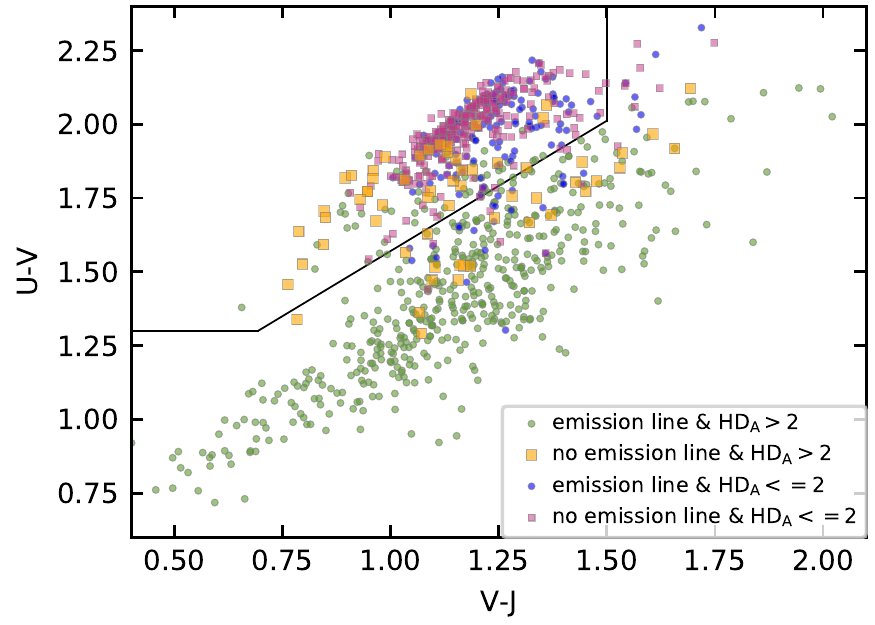}
  \includegraphics[width=0.49\textwidth]{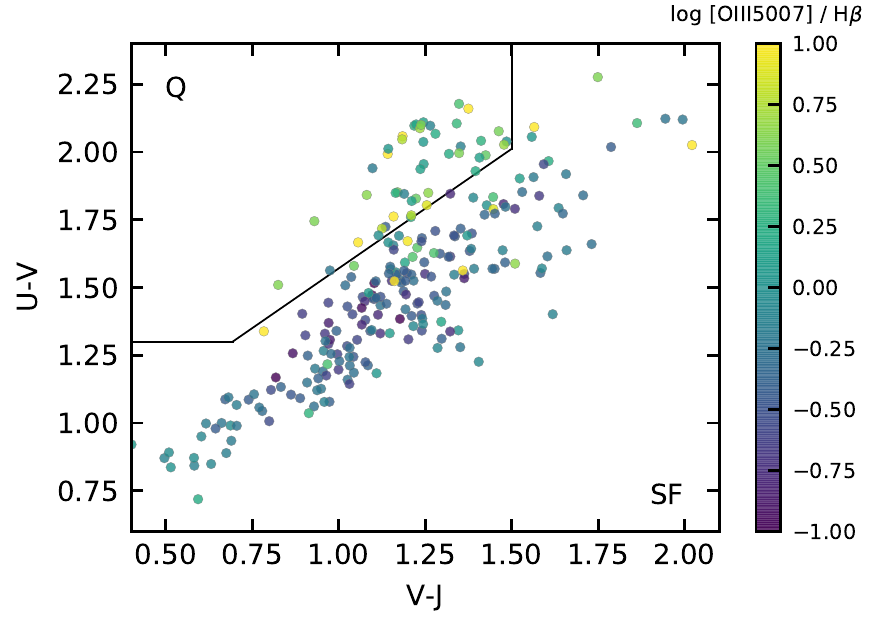}
  \caption{Left: UVJ diagram with emission line (non-)detections indicated. Emission line detected galaxies ($>5\sigma$) with \hda$>2\ \rA$ (green dots) and non-detected galaxies with \hda$\leq2\ \rA$ (pink squares) occupy the star-forming and quiescent locus, respectively, of the total population. Galaxies with emission lines and \hda$\leq2\ \rA$ (blue dots) as well as non-detected galaxies with \hda$>2\ \rA$ (orange squares) exhibit more scatter. The latter can be post-starburst galaxies or dust-obscured and star-forming, but in general tend to have more quiescent-like colors than most star-forming galaxies. The scatter of emission line detected \hda$\leq2\ \rA$ and non-detected \hda$>2\ \rA$ sources illustrates the complexity of galaxy evolution and suggests caution while classifying galaxies based on photometric UVJ colors. Right: UVJ diagram color-coded by [O{\sc III}]5007$/$\hb\ for sources selected by wavelength coverage, but including $<5\sigma$ detections in [O{\sc III}]5007 or \hb. Sources with log [O{\sc III}]5007$/$\hb$>0.5$ are mostly found within or close to the quiescent region of the diagram. We also detect slightly elevated ratios (log [O{\sc III}]5007$/$\hb$\sim0$) for very blue star-forming galaxies, indicating perhaps higher metallicity.}
  \label{fig:uvjem}
\end{figure*}

The left panel of Figure~\ref{fig:uvjem} shows which galaxies in the UVJ diagram have $>5\sigma$ detected emission lines (any of those in the catalog). Lowering the threshold does not significantly alter the results. The detections and non-detections are subdivided into populations with low ($\leq2\ \rA$) and high ($>2\ \rA$) \hda\ values{, based on the apparent bi-modal distribution in Figure \ref{fig:min4}}. Most of the galaxies with a {relatively} young stellar population (\hda$>2\ \rA$) have detected emission lines (green dots), consistent with their location in the UVJ diagram. In a similar manner, most galaxies with a {relatively} old population (\hda$\leq2$) located within the quiescent region have no emission lines detected (pink squares). However, there are also galaxies with old stellar populations, but significant line emission (blue dots). They tend to have similar $U-V$ and $V-J$ colors as quiescent galaxies, but exhibit more scatter. Finally, we identify galaxies with young stellar ages, but without any emission line detections (orange squares). Some of those are post-starburst galaxies. Others would be classified as dust-obscured and star-forming with red $V-J$. Most interestingly, their average locus is clearly offset from the general emission line detected star-forming population (green datapoints). This diversity of possible combinations of rest-frame colors, absorption line strengths, and emission line detections, {which has also been pointed out by \citet{Moresco13} using data from zCOSMOS, }suggests caution while classifying galaxies based on photometric $U-V$ and $V-J$ colors and warrants the necessity to obtain spectroscopic measurements to more fully understand galaxy evolution{ \citep[see also][]{Siudek18}}.

Finally, in the right panel of Figure~\ref{fig:uvjem} we show [O{\sc III}]5007$/$\hb\ in the context of the UVJ diagram for sources selected in the same way as for Figure~\ref{fig:min2}. We find excess ratios (log [O{\sc III}]5007$/$\hb$>0.5$) mostly either for quiescent galaxies, or star-forming galaxies with $U-V$ and $V-J$ colors close to those of quiescent galaxies. We also find that the bluest star-forming galaxies have relatively high ratios: log [O{\sc III}]5007$/$\hb$\sim0-0.25$, indicative of a lower metallicity. 

\subsection{\hda\ versus Fe4383}\label{sec:agemetal}

\begin{figure}
  \includegraphics[width=0.49\textwidth]{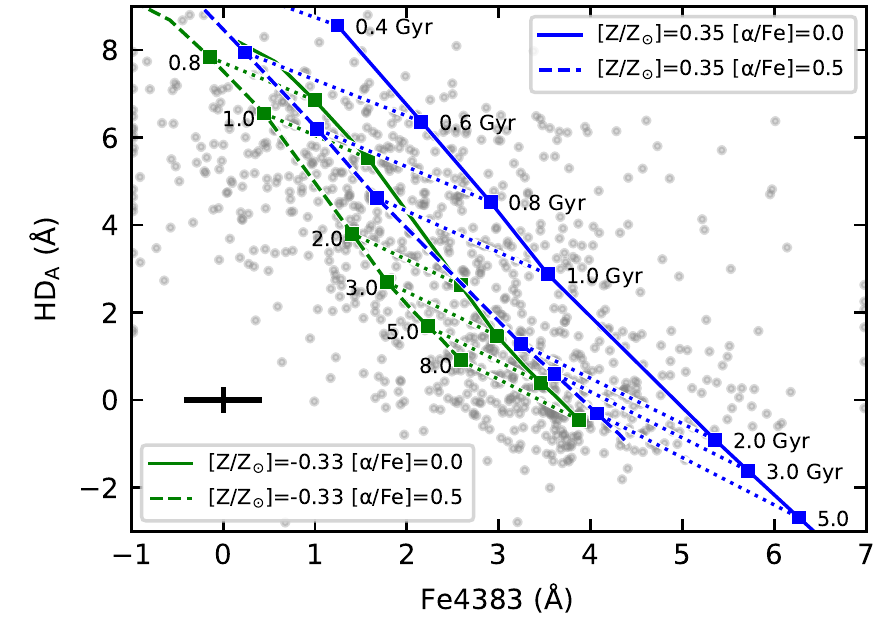}
  \caption{\hda\ versus Fe4383 for close to 1000 galaxies, for a varied sample ($2\lesssim$ \hda\ $(\rA)\lesssim 8$). A typical errorbar is shown in the bottom left. {Shown as well} are {the predicted index values from the single stellar population models of \citet{Thomas04}, assuming} subsolar {(green)} and supersolar {(blue)} metallicity {(solar metallicity predictions lie in between)}, low and high $\mathrm{[\alpha/Fe]}$, and spanning a range of stellar ages. This is the first time {age and metallicity absorption line indicators are available} with high $S/N$ {for such a large sample} at $z>0.6$. A full analysis of this data will be presented in a future paper.}
  \label{fig:femg}
\end{figure}

With Figure~\ref{fig:femg} we illustrate a different region of parameter space as probed by LEGA-C, {whose high-resolution, high$-S/N$ spectra allow us to measure a variety of stellar absorption features with distinct sensitivity to age, metallicity, and element abundance ratios.} We have Fe4383 measurements for 921 ($S/N (4000\rA)>10\ \mathrm{pix^{-1}}$) galaxies with \hda\ ranging from $-2\ \rA$ to $8\ \rA$. {We also show \hda\ and Fe4383 predictions from the models by \citet{Thomas03,Thomas04}, which are based on the evolutionary population synthesis of \citet{Maraston98} of single stellar populations with a Salpeter IMF. The galaxies in LEGA-C} span {at least} the full range of parameters, from subsolar ($\mathrm{[Z/Z_{\sun}]}=-0.33$) to supersolar ($\mathrm{[Z/Z_{\sun}]}=0.35)$ metallicity, $\mathrm{[\alpha/Fe]}=0$ to $\mathrm{[\alpha/Fe]}=0.5$, and ages ranging 0.4 to {8} Gyr. {By interpreting optimal sets of absorption features we will be able to reduce the degeneracies between stellar population physical parameters \citep[e.g.,][]{Worthey94,Gallazzi05,SanchezBlazquez09,SanchezBlazquez11}.} A full-fledged analysis of metallicities, ages, and corresponding scaling relations of galaxies in LEGA-C will be presented in a future paper.

\section{Summary}

In this paper we presented Data Release II of the LEGA-C survey, comprising of a total of \totn\ spectra (\prims\ primary target and \fills\ fillers) in the COSMOS field with a typical continuum $S/N\simeq20\ \rA^{-1}$ as well as a catalog with stellar and gas observed velocity dispersions, Lick/IDS indices, and emission line strengths and equivalent widths. The spectroscopic redshifts are unambiguous due to the many absorption features in the spectra and are only flagged if they could not be obtained at all. {Published} uncertainties include both formal errors and additional systematic uncertainties facilitated by an analysis of duplicated observations.

We show that for the first time at significant lookback time {($\sim8$ Gyr)} we can explore the parameter space connecting dynamical and stellar population properties for a large sample ($>1000$) of galaxies. We find that \hda\ is bi{-}modally distributed and follows a broad trend with observed (spatially integrated) stellar velocity dispersion, such that higher dispersion galaxies tend to have lower \hda\ values, indicative of higher stellar ages (or lower metallicities). Galaxies with low \hda\ and higher observed velocity dispersions are not necessarily all quiescent: in addition we find instances of both old galaxies with emission lines and young galaxies without emission lines. Such complexity is even more apparent if these properties are explored in the context of the UVJ diagram, again revealing a diversity of possible combinations of rest-frame colors, absorption line strenghts and emission line detections, suggesting caution while classifying galaxies based on photometric UVJ colors alone. For $\sim1000$ galaxies we present measurements of different age/metallicity indicators and show that these span (and may constrain) a large range of stellar population models.

Data aquisition of LEGA-C was completed in March 2018. Upcoming data releases are planned for late 2018 / early 2019 and will additionally include model-based stellar population properties.

\acknowledgements{}
Based on observations made with ESO Telescopes at the La Silla Paranal Observatory under programme ID 194-A.2005 (The LEGA-C Public Spectroscopy Survey). This project has received funding from the European Research Council (ERC) under the European Union’s Horizon 2020 research and innovation programme (grant agreement No. 683184). KN and CS acknowledge support from the Deutsche Forschungsemeinschaft (GZ: WE 4755/4-1). We gratefully acknowledge the NWO Spinoza grant. JvdS is funded under Bland-Hawthorn's ARC Laureate Fellowship (FL140100278)

\appendix
\section{Example LEGA-C spectra}

\begin{figure*}
  \includegraphics[width=\textwidth]{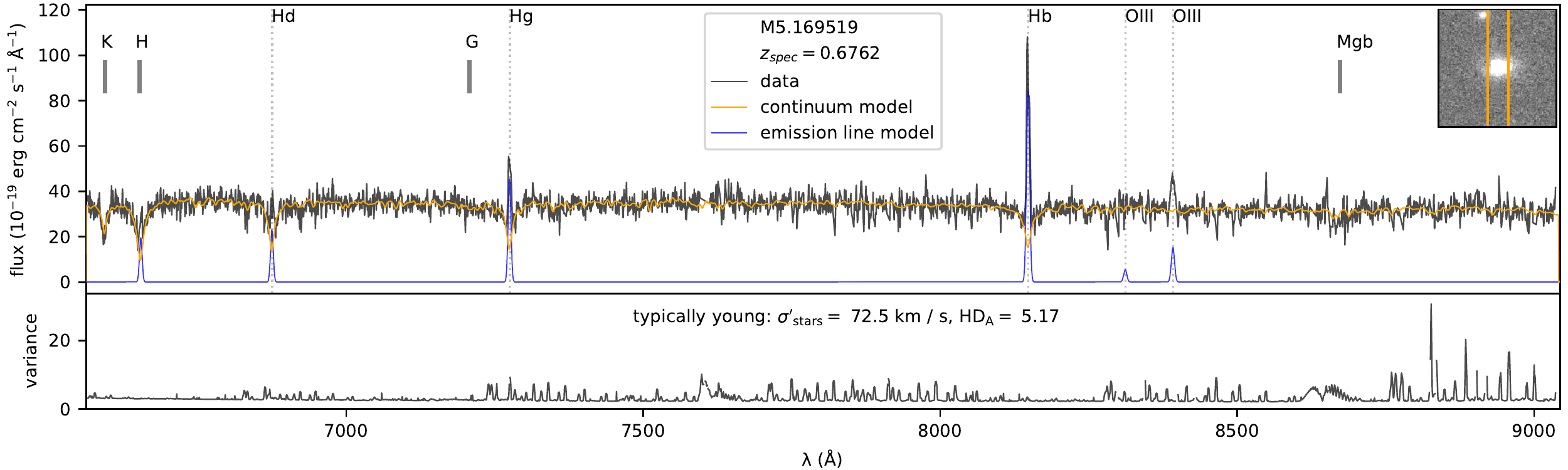}\vspace{8pt}
  \includegraphics[width=\textwidth]{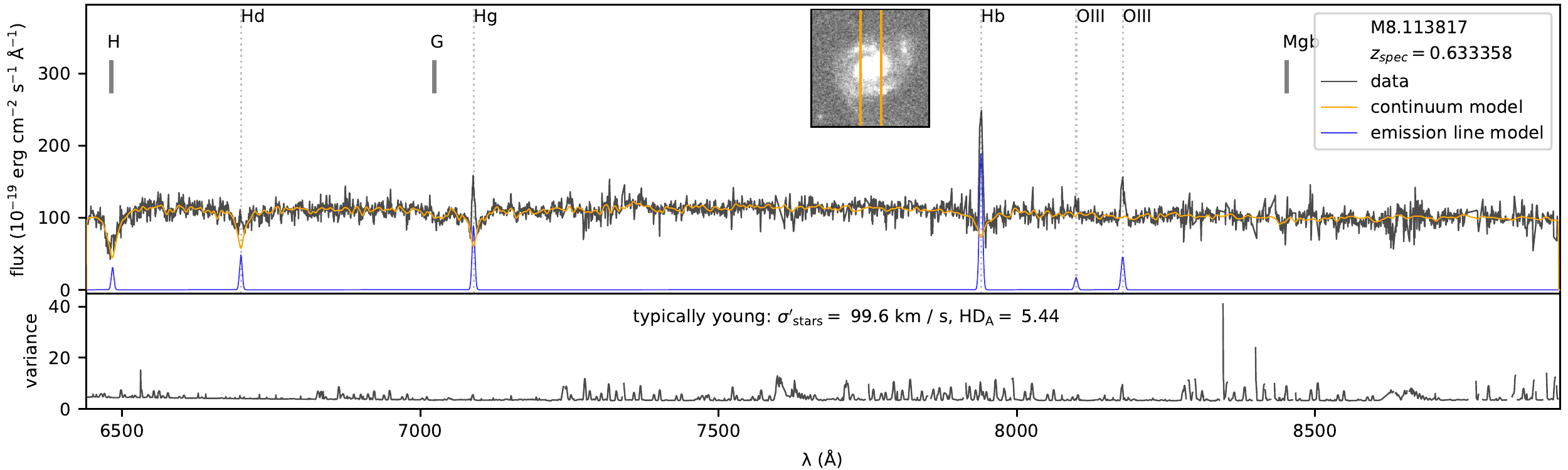}\vspace{8pt}
  \includegraphics[width=\textwidth]{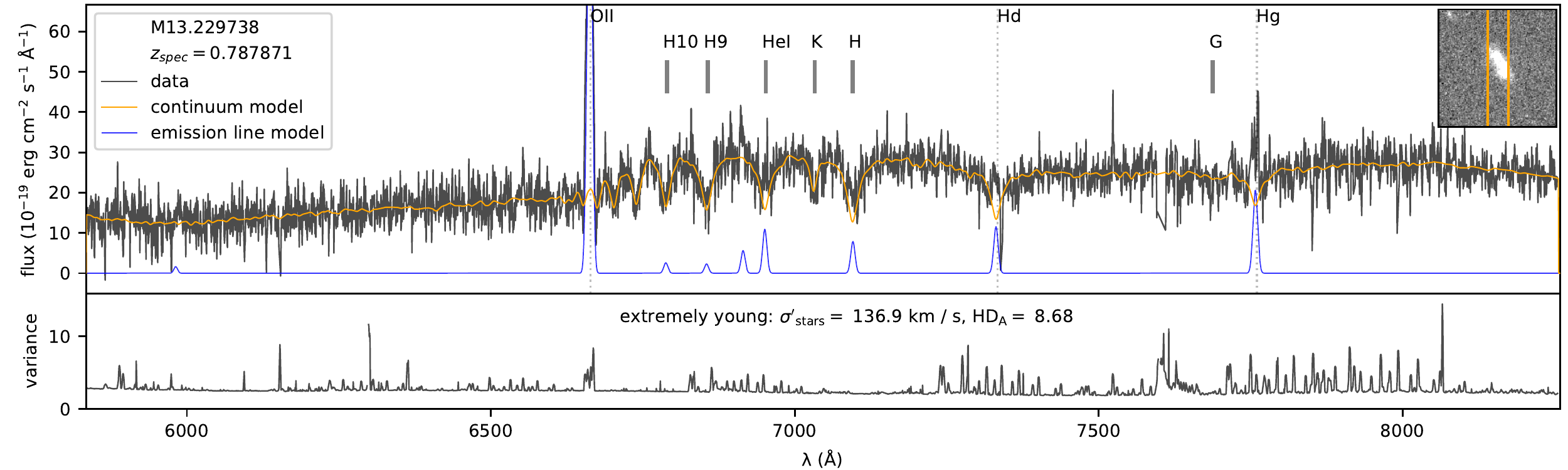}
  
   \caption{Examples of galaxies occupying different regions in $\sigma'-$\hda\ parameter space. Shown are the flux and variance in the same units, as well as the best-fit continuum and emission line models from pPXF. Gray lines indicate important emission or absorption lines. Insets are $5.4\arcsec\times5.4\arcsec$ HST/ACS images with the slit dimensions in the North-South direction indicated by vertical lines.}
  \label{fig:spectra}
\end{figure*}

\addtocounter{figure}{-1}
\renewcommand{\thefigure}{\arabic{figure} -- continued}

\begin{figure*}
  \includegraphics[width=\textwidth]{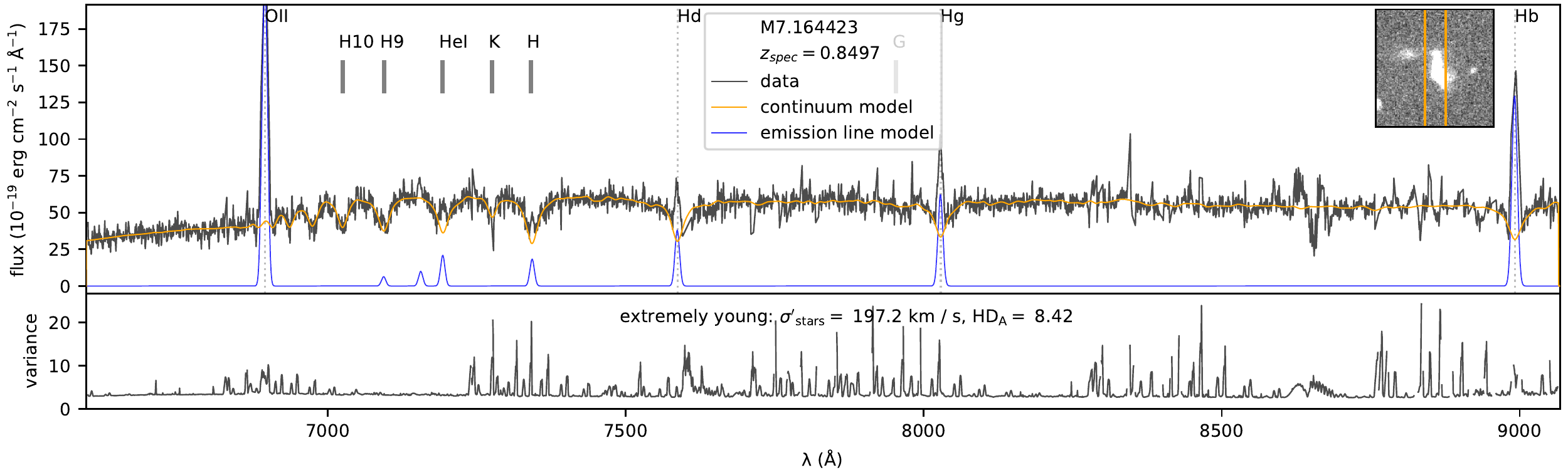}\vspace{8pt}
  \includegraphics[width=\textwidth]{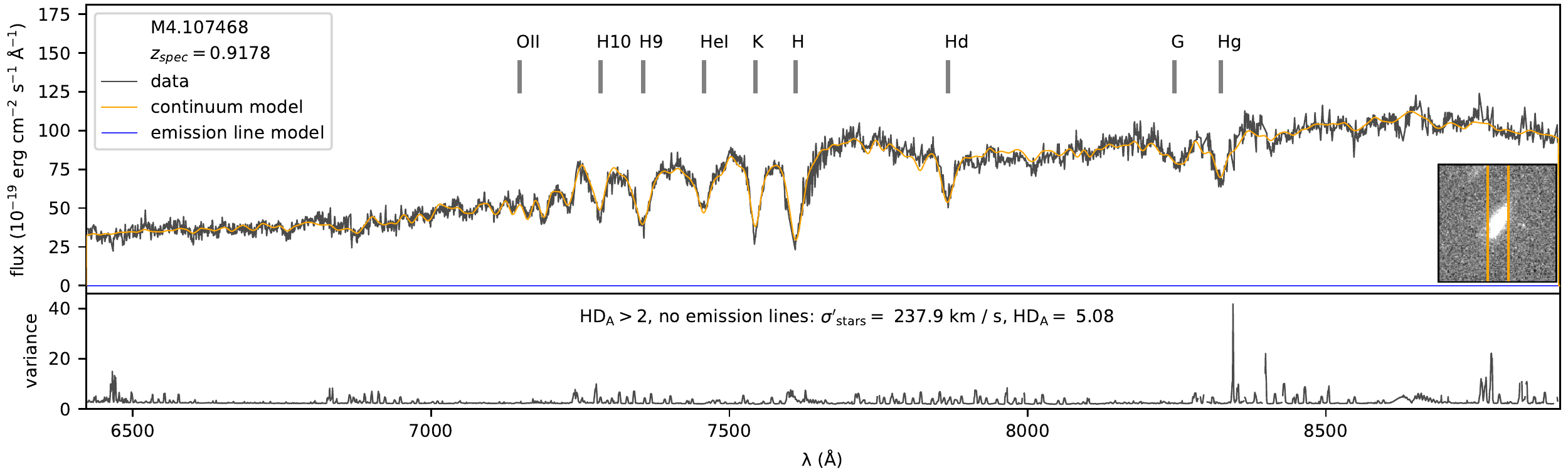}\vspace{8pt}
  \includegraphics[width=\textwidth]{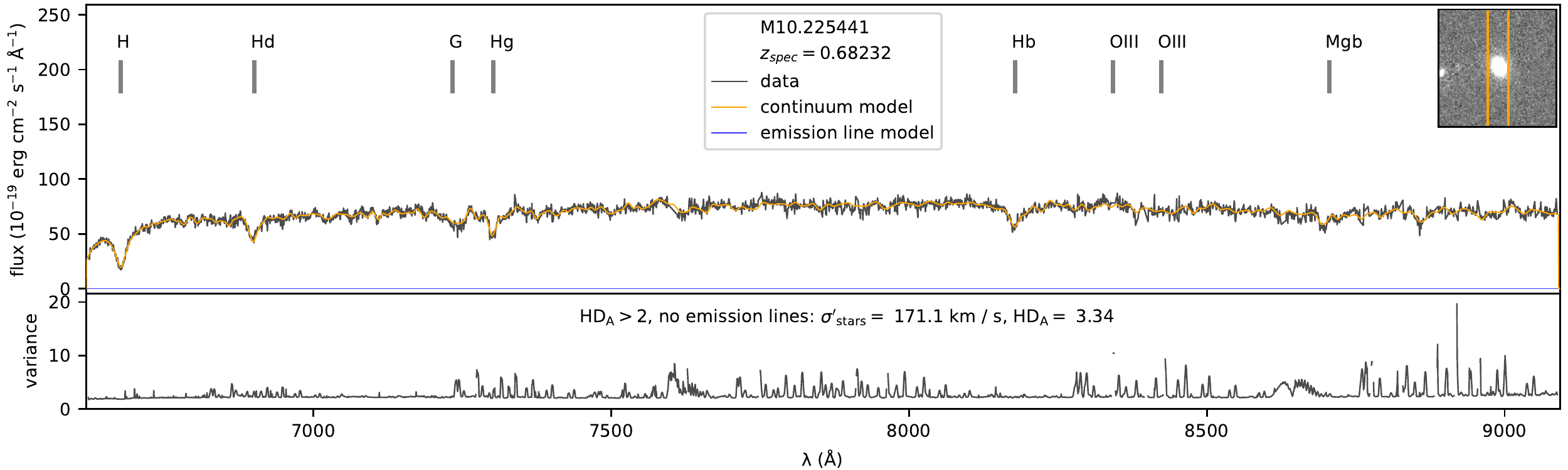}\vspace{8pt}
  \includegraphics[width=\textwidth]{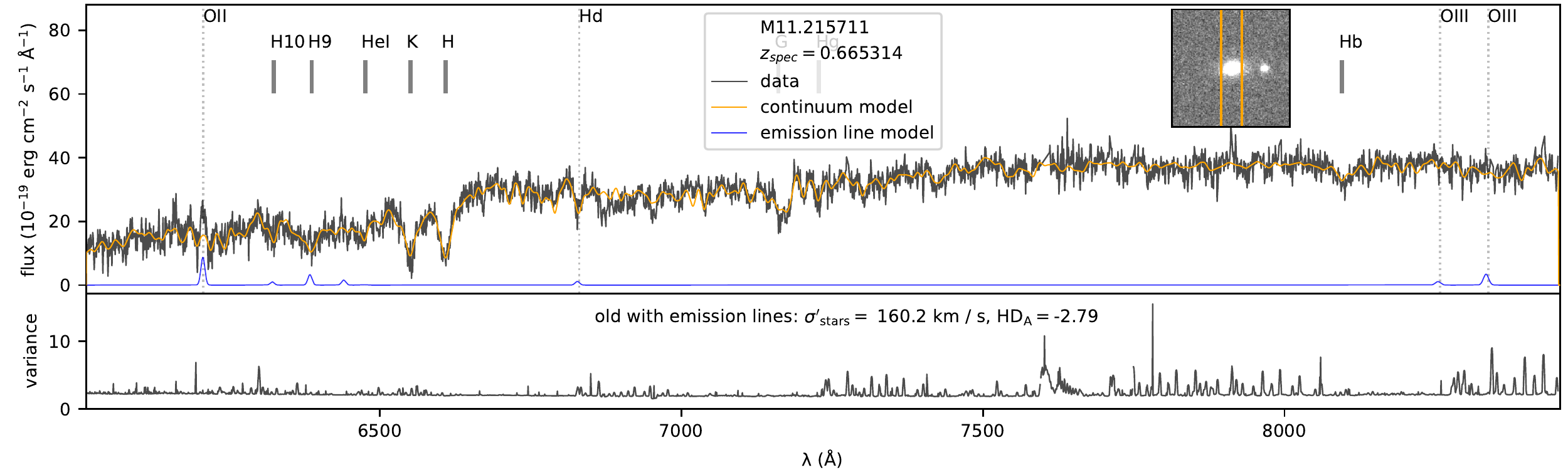}
  
  \end{figure*}

\addtocounter{figure}{-1}

\begin{figure*}
  \includegraphics[width=\textwidth]{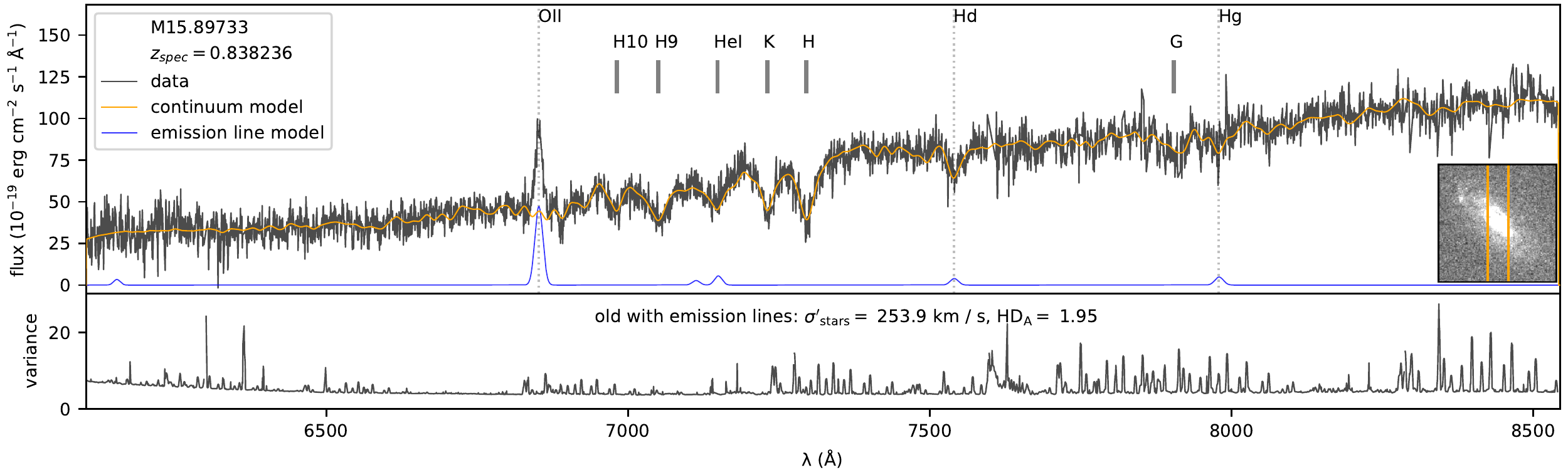}\vspace{8pt}
  \includegraphics[width=\textwidth]{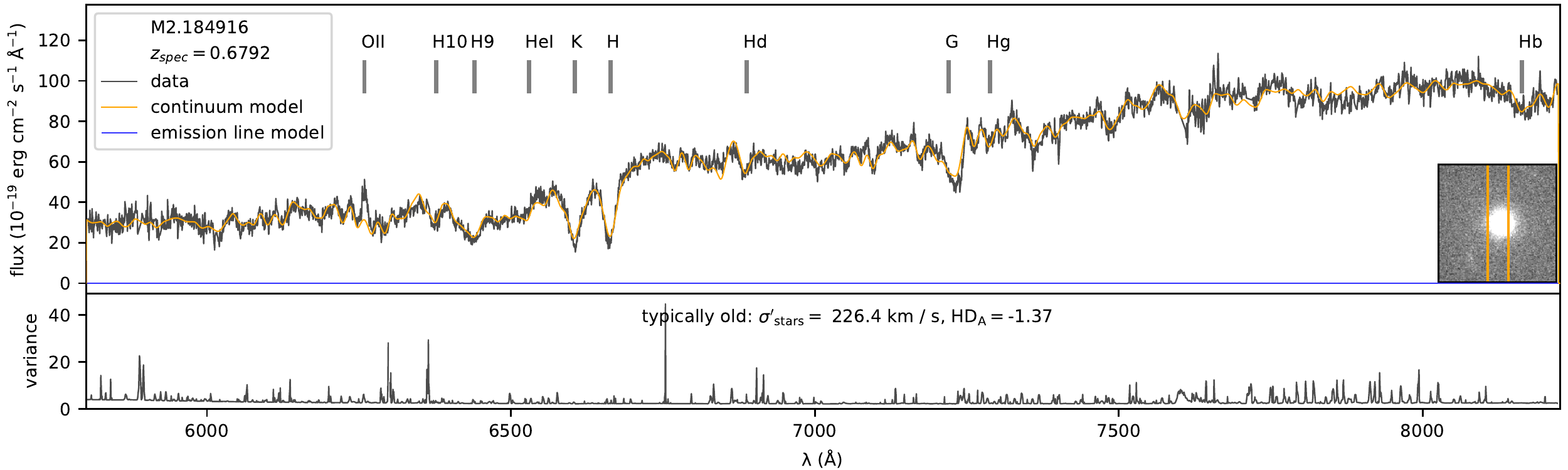}\vspace{8pt}
  \includegraphics[width=\textwidth]{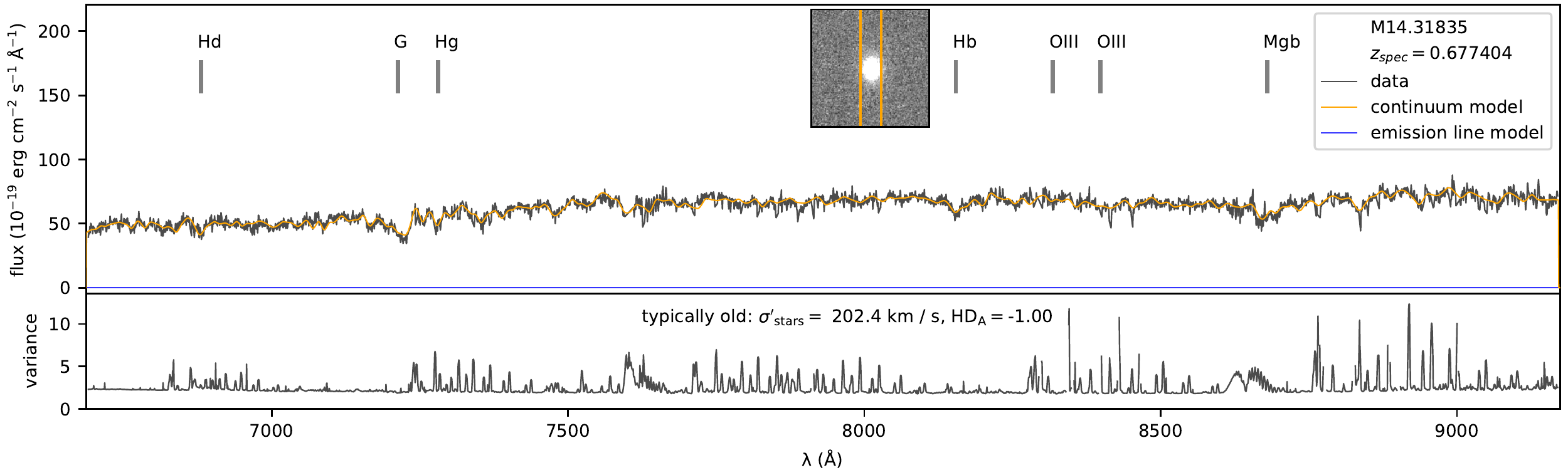}
\end{figure*}

In Section~\ref{sec:science} we discussed the complexity of galaxy evolution as illustrated by the great variety in spectra in terms of observed velocity dispersions, \hda\ absorption strengths, and emission line detections. In Figure~\ref{fig:spectra} we show examples of typically young (\hda$>2$), extremely young (\hda$>8$), and typically old (\hda$\leq2$) galaxies, as well as young galaxies without emission line detections or old galaxies with emission lines.

\end{document}